\documentclass[twocolumn,pra,amsmath,amssymb,floatfix,superscriptaddress]{revtex4}

\usepackage{graphicx}
\usepackage{epstopdf}
\usepackage{dcolumn}
\usepackage{bm}
\usepackage{amsmath}
\usepackage{amssymb}
\usepackage{setspace}
\usepackage{hyperref}
\usepackage{color}
\hypersetup{%
  pdfpagemode=None, 
  pdfstartpage=1,
  pdfstartview=FitH,
  pdfmenubar=true,
  pdftoolbar=true,
  colorlinks = true,
  linkcolor=blue,
  citecolor=blue,
  bookmarksopen=false
}

\definecolor{green}{rgb}{0,.4,0}

\renewcommand{\equationautorefname}{Eq.}
\def\equationautorefname~#1\null{Eq. (#1)\null}
\newcommand{\ie}{i.\,e.\,}%
\begin{document}

\title{Defining the $p$-wave scattering volume in the presence of dipolar interactions}

\author{Anne Crubellier}
\email{anne.crubellier@u-psud.fr}
\affiliation{Laboratoire Aim\'e Cotton, CNRS, Universit\'e Paris-Sud,
  Universit\'e Paris-Saclay, ENS Cachan, Facult\'e des Science
  B\^atiment 505, 91405  Orsay Cedex, France} 

\author{Rosario Gonz\'alez-F\'erez}
\email{rogonzal@ugr.es}
\affiliation{Instituto Carlos I de F\'{\i}sica
  Te\'orica y Computacional and Departamento de F\'{\i}sica At\'omica,
  Molecular y Nuclear, Universidad de Granada, 18071 Granada,
  Spain}

\author{Christiane P. Koch}
\email{christiane.koch@uni-kassel.de}
\affiliation{Theoretische Physik,  Universit\"at Kassel,
  Heinrich-Plett-Str. 40, 34132 Kassel, Germany}

\author{Eliane Luc-Koenig}
\email{eliane.luc@u-psud.fr}
\affiliation{Laboratoire Aim\'e Cotton, CNRS, Universit\'e Paris-Sud,
  Universit\'e Paris-Saclay, ENS Cachan, Facult\'e des Science
  B\^atiment 505, 91405  Orsay Cedex, France} 

\date{\today}
\begin{abstract}
The definition of the scattering volume for $p$-wave collisions needs to be generalized in the presence of dipolar interactions for which the potential decreases with the interparticle separation as $1/R^3$. Here, we propose 
a generalized definition of the scattering volume characterizing the short-range interactions in odd-parity waves, 
obtained from an analysis of the $p$-wave component of the two-body threshold wave function. Our approach uses an 
asymptotic model and introduces explicitly the anisotropic dipole-dipole interaction, which governs the ultracold collision 
dynamics at long-range. The short-range interactions, which are essential to describe threshold resonances, are taken into 
account by a single parameter which is determined by the field-free $s$-wave scattering length.
\end{abstract}
\maketitle
%
\section{Introduction  }
\label{sec:intro}
Collisions in ultracold atomic or molecular gases are universally described by the $s$-wave scattering length in case of bosons and unpolarized fermions or by the $p$-wave scattering volume in case of spin-polarized fermions. The value determines the strength of the interaction, which is repulsive (attractive) if the scattering length is positive (negative)~\cite{Dalibard98}.  
Experimental control of the scattering properties, both scattering length and scattering volume, is a long-standing goal in ultracold gases~\cite{GiorginiRMP08,ChinRMP10,TheisPRL04,BlattPRL11,YamazakiPRA13,KillianPRL13}. 

Given the prominence of the scattering parameters, it is somewhat 
unsatisfactory that  all scattering parameters, even the scattering length, cannot be defined for an isotropic potential decreasing asymptotically as $1/R^3$~\cite{OMalleyJMP61,Joachain83}.
In detail, the tangent of the scattering phase shift at low energy cannot be expanded in powers of the wave number $k$ of the incident wave.
Simultaneously, the asymptotic threshold wave function includes an $\ln(R)$ contribution in addition to the $(R-a)$ term that defines the scattering length $a$. 
In contrast, for an anisotropic interaction decreasing as $1/R^3$, 
the $s$-wave scattering length is unambiguously defined. In this case, the
effective $s$-wave  potential decreases more rapidly, as $1/R^4$, which results in a 'quasi-long range' 
character of the dipole-dipole interaction~\cite{MarinescuPRL98,DebPRA01}. In a previous study of non-resonant light 
control~\cite{CrubellierPRA17}, we have verified this assertion by analyzing a particular threshold solution, the one that 
asymptotically decreases in all $\ell >0$ channels while linearly increasing in the $\ell=0$ channel. Here, we examine 
the definition of a generalized $p$-wave scattering volume for an anisotropic $1/R^3$ interaction, which is
both an open problem and a prerequisite for studying non-resonant light control of $p$-wave scattering~\cite{Crubellier18b}.

The paper is organized as follows. In~\autoref{sec:scatt-vol-Levy-Keller}, we define the generalized scattering volume for 
two interacting ultracold atoms in the presence of a non-resonant laser field. The field-dressed generalized scattering 
volume is numerically determined in~\autoref{sec:numerical}, and we show that it presents a divergence 
each time a bound state appears at threshold. 
We conclude in~\autoref{sec:conclusion}. 

%
\section{Generalized scattering volume in the presence of $R^{-3}$ interaction}
\label{sec:scatt-vol-Levy-Keller}
%
We start by formulating the problem in Sec.~\ref{subsec:kx-LK} and then employ
an extension of the single-channel Levy-Keller approach~\cite{LevyJMP63} (described in appendix~\ref{app:sec-Levy-Keller}) to determine the scattering wave function needed to define the scattering volume. 
The Levy-Keller approach 
considers a
multipolar expansion of the effective $p$-wave potential and introduces pairs of analytical reference functions. The scattering wave function is written as a combination of these two reference functions and 
one focuses, as usual, on the ratio $\mathcal M(R)$ between the amplitude of these two functions. 
Above the dissociation limit, when the spherical Bessel functions are used as reference functions, $\mathcal M(R)$ is 
identical to the tangent of the local $p$-wave phase of scattering theory. 

In \autoref{subsec:kx-infty}, we consider the zero energy limit of $\mathcal M(R)$ and show that
in a potential decreasing  asymptotically as $-1/R^3$, $\mathcal M(R)$, for
 very large $R$, 
becomes proportional to the wave number $k$ (not to $k^3$). This prevents the definition of the scattering 
volume. 
However, information on the short-range interaction
which is what is captured by $\mathcal{M}(R)$
can be obtained by restricting the $R$-range so that $kR$ becomes not too large. 

We show, in~\autoref{subsec:kx-eq-0}, that this is equivalent to analyzing the asymptotic behavior of the threshold wave function using a pair of functions with asymptotic form $\sim R^2$ and $\sim 1/R$, a method frequently used in scattering theory. Then, the analytical expression of $\mathcal M(R)$ contains, besides divergent asymptotic terms involving the multipolar parameters of the asymptotic potential, a constant term $\mathcal{M}_0$ depending on the short-range interactions only. This term $\mathcal{M}_0$ is identified as the generalized scattering volume. 
%
\subsection{Statement of the problem}
\label{subsec:kx-LK}
%
The asymptotic Schr\"odinger equation, in the Born-Oppenheimer approximation, describing the nuclear relative motion of a pair of atoms interacting with a non-resonant laser field reads 
\begin{equation}
  \label{eq:asy}
  \left[-\frac{d^2}{dx^2}+ \frac{\mathbf{L}^2}{x^2}  - \frac{1}{x^6} 
    - {\mathcal I}  \frac{\cos^2\theta -1/3}{x^3} - {\mathcal E}
  \right] f (x,\theta,\phi) = 0\,,
\end{equation}
where reduced units (ru) of length $\sigma$, energy $\epsilon$ and light intensity $\beta$~\cite{CrubellierPRA17}, defined in appendix~\ref{app:red-units},
 have been used. The first two terms stand  for the vibrational and rotational kinetic energies, $f(x,\theta,\phi)$ is the 
 asymptotic wave function, $x$ the interparticle separation, $(\theta,\phi)$ the Euler angles, and ${\mathcal I}$ the 
 non-resonant laser intensity. 
The third term is  the van der Waals interaction, described by the universal term $-1/x^6$. 
 The  fourth term in~\autoref{eq:asy} 
 corresponds to dipole-dipole interaction, either for permanent dipoles such as found in the scattering between polar molecules or for an induced dipole coupling to the (non-resonant) field~\cite{Crubellier18b}.
The non-resonant field intensity ${\mathcal I}$ in ru is a tunable parameter allowing for the control of the collision.
This has been discussed for even-parity $\ell$ states and $m=0$, providing a means to tune the $s$-wave scattering length~\cite{CrubellierPRA17}. 
Here, we consider collisions in odd-parity states with $m=0$ or $m=\pm 1$.

In scattering theory, a first method to determine the scattering parameter in the channel $\ell$ consists in analyzing the 
asymptotic form of the physical solution of the Schr\"odinger equation in this channel, at a vanishingly small positive 
energy. This physical solution is constructed as a superposition of  the regular $kx\,j_\ell(kx)$ and  irregular  
$kx\,\eta_\ell(kx)$ spherical Bessel functions, where the latter is multiplied by $-\tan \delta_\ell(k,x)$, with $\delta_\ell(k,x)$ 
being  the local phase converging to the asymptotic phase shift, and $k$ the wave number. The scattering parameter 
$(a_\ell)^{2\ell+1}$, which has the dimension of a length to the power of $(2\ell+1)$ and characterizes elastic collisions, is defined by the following low-energy limit 
\begin{equation}
  \label{eq:def_a_ell}
  \lim_{k \rightarrow 0}  -\frac{\tan \delta_\ell(k)}{k^{2\ell+1}} =
  \frac{(a_\ell)^{2\ell+1}}{(2\ell+1)!! (2\ell-1)!! }\,,
\end{equation}
where $\delta_\ell(k)$ is the asymptotic phase shift. Notice that this limit~\eqref{eq:def_a_ell} does not exist for a potential decreasing as $-1/x^q$ with $2\ell+3 \ge q$~\cite{Landau,OMalleyJMP61}, because the tangent of the asymptotic phase shift increases  as $k^{q-2}$, independently of $\ell$. Thus, the scattering volume for $\ell=1$, \ie, $\mathcal V=(a_{\ell=1})^3/3$ (a factor of $3$ is included to simplify further notation) is defined only for a potential decreasing asymptotically at least as $-1/x^6$~\cite{Joachain83}. As a consequence, the scattering volume is not defined for the $-1/x^3$ potential appearing in the asymptotic Schr\"odinger equation~\eqref{eq:asy}. 

A second method to calculate the scattering parameters $(a_\ell)^{2\ell+1}$ writes the asymptotic form of the zero energy 
wave function as a combination of the field-free regular $x^{\ell+1}$ and irregular $1/x^\ell$ solutions, the latter with a 
coefficient proportional to $-(a_\ell)^{2\ell+1}$. Thus, in spite of the slow decrease of the $-1/x^3$ potential, a generalized 
scattering volume, which characterizes the interactions at short range, can be defined in the limit $k \rightarrow 0$ and 
$x \rightarrow \infty$ 
when keeping $xk$ finite. This is the strategy that we will use in the following.

We start by inspecting the analytical solution of an approximation of the asymptotic Schr\"odinger equation~\eqref{eq:asy} 
for $k\ge 0$. We first expand the angular part of the wave function $f(x,\theta,\phi)$ in~\autoref{eq:asy} in spherical 
harmonics, each  multiplied by a radial wave function $u_\ell(x)$. 
In a single-channel approximation, we take into account the coupling between the different partial waves by introducing an 
effective $\ell$-dependent potential $V_\ell(x)$ given by its asymptotic multipolar expansion 
\begin{equation}
\label{eq:pot}
V_\ell(x)= -\frac{c_3^\ell}{x^3} -\frac{c_4^\ell}{x^4} -\frac{c_5^\ell}{x^5} -\frac{c_6^\ell}{x^6}\,.
\end{equation}
For each partial $\ell$-wave and energy $\epsilon=k^2\geq 0$, the radial Schr\"odinger equation 
\begin{equation}
\label{eq:Schr}
u_\ell''(x)-\left(\frac{\ell(\ell+1)}{x^2}+V_\ell(x)+k^2\right)\,u_\ell(x)=0
\end{equation}
can be solved analytically either exactly or using perturbation theory. For simplicity, the $\ell$-dependence of $V_\ell(x)$, 
the coefficients $c_i^\ell$,  and $u_\ell(x)$ is omitted  from now on. 

The two-potential  Levy-Keller method~\cite{LevyJMP63,HinckelmannPRA71}, which is described  in appendix~\ref{app:sec-Levy-Keller}, constructs the solution of the radial Schr\"odinger equation~\eqref{eq:Schr} as the following linear combination of two reference functions $(\varphi(x),\psi(x))$
\begin{equation}
\label{eq:fctLK}
u(x)={\mathcal A}(x)\Big(\varphi(x)-\psi(x)\,{\mathcal M}(x) \Big). 
\end{equation}
These reference functions $(\varphi(x),\psi(x))$ are solutions of the Schr\"odinger equation~\eqref{eq:Schr} at the same energy but for the potential $V_f(x)=-c_{pf}/x^q$. If the dominant term in $V(x)$ is $-c_{pf}/x^q$, $V_f(x)$ is a zeroth-order approximation to $V(x)$ and, then, $u(x)$ is the zeroth-order solution of~\autoref{eq:Schr}. The function $\mathcal M(x)$ in~\autoref{eq:fctLK}  satisfies the non-linear first-order differential equation~\eqref{eq:solLK-M}, and the logarithmic derivative of $\mathcal A(x)$ the first-order differential equation~\eqref{eq:solLK-A} involving $\mathcal M(x)$. Each of these differential equations~\eqref{eq:solLK-M} and~\eqref{eq:solLK-A} introduces a single integration constant $\mathcal M_0$ and $\mathcal A_0$, respectively. $\mathcal A_0$ is a global multiplicative constant without physical meaning, whereas $\mathcal M_0$ is an additive constant, which can be identified as the generalized scattering volume, as shown below. This is in line with $\mathcal M(x)$ in $u(x)$~\autoref{eq:fctLK} taking the role of the tangent of a local phase, apart from the fact that it does not converge as $x$ increases. 

In this work, we use three potentials $V_{f}(x)$ as zeroth-order approximation to $V(x)$. The corresponding reference 
pairs $(\varphi(x), \psi(x))$ are labeled as BC$2k$ for $k>0$ and BC2 or BC23 for $k=0$, and are presented 
in~\autoref{tab:fct} of appendix~\ref{app:subsec-basis-function}. For $\ell=1$, the BC$2k$, BC2 and BC23 reference 
functions $\varphi(x)$ and $\psi(x)$ vary as $x^2$ and $1/x$ for not too high  values of $kx$, everywhere, and for 
$x \rightarrow \infty$, respectively. In all three cases, $\mathcal M(x)$ represents (in reduced units) a quantity that has 
dimension of volume. 

Once the solution of the approximate single-channel Schr\"odinger equation~\eqref{eq:Schr} is constructed for $\ell=1$ and $k>0$,  the limit of $\mathcal M(x)$ for $k\to 0_+$ is evaluated. 
From this limit of $\mathcal M(x)$, we extract the constant term $\mathcal M_0$ suitable to define the (approximate) generalized scattering volume $\mathcal V$, cf.~\autoref{eq:def_a_ell}.
%
\subsection{Evaluating the $k\to 0_+$ limits ($kx\rightarrow \infty$ and $kx$ small)}
\label{subsec:kx-infty}
%
The scattering length and scattering volume are defined at very low positive energy $k\rightarrow 0_+$ and in the limit $x\to\infty$, see~\autoref{eq:def_a_ell}. We first show that this standard way  cannot be used to define the scattering volume in the presence of the dipole-dipole interaction (contrary to the scattering length~\cite{CrubellierPRA17}). 
The contribution of the long-range part of the potential $x\ge d$ to the tangent of the asymptotic phase shift $\tan \delta_{\ell=1}(k)=\lim_{x \rightarrow \infty} \mathcal M(x)$, can be expressed in terms of the tangent of the short-range phase at $x=d$, $t_0=\tan [\delta_{\ell=1}(k,x=d) ]$, 
which accounts for all of the short-range ($x<d$) contributions of the potential to $\mathcal M(x)$ and of the integrals
$$\mathcal J_{\Phi,\Psi}(d) = k\,{c_3} \lim_{x \rightarrow \infty} \int_{kd}^{kx} \frac{ \Phi(\rho) \Psi(\rho)}{ \rho^3} d\rho \, ,$$ 
occurring in  a treatment  to first order of the perturbation theory of~\autoref{eq:solLK-M}. Here
 $\Phi(\rho)$ and $\Psi(\rho)$ denote one function $\varphi(x)$ or $\psi(x)$ of the reference pair BC$2k$ for $\ell=1$  given in~\autoref{tab:fct}. These integrals converge and are analytically known (see Refs.~\cite{HinckelmannPRA71,ShakeshaftJPB72}). The lower boundaries introduce terms proportional to $k\times (kd)^q $ with $q>-4$, whereas the upper ones converge proportionally to $k/(kx)^2$ 
(see~\autoref{eq:phase-infty-a}). The tangent of the asymptotic phase shift reads
\begin{subequations}
\label{eq:phase-infty}
\begin{eqnarray}
  \tan \delta_{\ell=1}(k) &=&  c_3 \,\bigg(\frac{k}{4} - \frac{  d^2 k^3}{18}\bigg) 
  +t_0 \,\bigg(1  +\frac{2 c_3 }{3d} -\frac{4 c_3 d k^2}{15} \bigg)\nonumber\\
                          &&+t_0^2\, c_3\,\bigg( \frac{1  }{4 d^4 k^3} +  \frac{1 }{2 d^2 k}  - \frac{k}{4}  + \frac{d^2 k^3 }{18  } \bigg) \, \cdots  .
                             \label{eq:phase-infty-a}
\end{eqnarray}
Restricting $t_0$ to its lowest term with the usual $k^3$ dependence, $t_0\sim-A\,k^3$, results  at low-energy in
\begin{eqnarray}
\tan \delta_{\ell=1}(k)\sim  k\,\frac{c_3}{4}   
 - k^3\,\bigg(A - \frac{c_3 A^2}{4  d^4} + \frac{2 A c_3}{3d}   + \frac{c_3  d^2}{18}\bigg) \, \cdots . 
\label{eq:phase-infty-b}
\end{eqnarray}
\end{subequations}
This expression is identical to that of Ref.~\cite{ShakeshaftJPB72} using the procedure introduced in 
Ref.~\cite{HinckelmannPRA71} and disregarding the effective range contribution to $t_0$.
The first term in $\tan \delta_{\ell=1}(k)/k^3$ obviously  diverges for $k \rightarrow 0_+$ and the scattering volume cannot 
be defined. Note that this divergence is entirely determined by the asymptotic part of the potential $c_3$, and has a 
universal character since it is independent of the short-range interaction. This is related to the quasi-universal character of 
dipolar scattering in cold and ultra-cold gases governed by the potential barrier for $\ell \ge 1$~\cite{BohnNJP09}. 
The convergence toward the asymptotic value given in~\autoref{eq:phase-infty} results from the $x$-dependent term $k c_3 (1+t_0^2)/4k^2x^2$ in the upper boundary of the integrals $\mathcal J_{\Phi,\Psi}(d)$
 that decreases slowly. 
This motivates a second procedure to evaluate the threshold limit by taking $k\to 0_+$ and $x \to\infty$ but keeping $kx$ finite. In this limit, the influence of the inner part of the potential becomes relevant. This treatment captures the physical features of experiments trapping particles at ultralow temperatures. 

%
\subsection{Defining the scattering volume via the threshold wave function}
\label{subsec:kx-eq-0}
%
For a given long-range potential $V(x)$ in~\autoref{eq:pot}, and a pair of reference functions defined from the single-term potential 
$V_f(x)$, ${\mathcal M}(x)$ satisfies the Ricatti equation~\eqref{eq:solLK-M}. 
At threshold ($k=0$) or just above it ($k \rightarrow 0_+$) and for intermediate values of $x$, such that 
the asymptotic form $V(x)\approx-c_3/x^3$ is already reached and simultaneously $kx$ remains very small, the BC$2k$ reference functions of $\ell=1$ can be replaced by the leading terms $\varphi(x)\approx(kx)^2/3$ and $\psi(x)\approx1/(kx)$, see~\autoref{tab:fct}. Inserting this approximation into the differential equation~\eqref{eq:solLK-M} for $\mathcal M(x)$, and setting ${\overline{\mathcal M}}(x)=3 {\mathcal M}(x)/k^3$, we obtain 
 $$\frac{d{\overline{\mathcal M}(x)}}{dx}=-\frac{c_3}{3\, x^3} \left(x^2- \frac{ {\overline{\mathcal M}(x)}}{x}\right)\,.$$ 
For the BC2 reference pair of the field-free Hamiltonian and $\ell=1$, \ie, $\varphi_{BC2}(x)=x^2$ and $\psi_{BC2}(x)=1/x$, 
${\mathcal M}(x)$ satisfies the same equation. Thus, in the limited $x$-range where $kx$ remains small, the relative amplitudes  
${\mathcal M}(x)$ obtained with the BC2 and BC$2k$ pairs at threshold and just above it 
are identical. In this work, we characterize the low-energy $p$-wave scattering  by the threshold wave functions at $k=0$, \ie, $\mathcal M(x)$ and $\mathcal A(x)$ are determined for the BC2 reference pair.

The solution of~\autoref{eq:solLK-M} is obtained by expanding ${\mathcal M}(x)$ into terms $1/x^q$ (with $q \ge -2$) and 
$\ln(x)/x^q$ with ($q\ge 0)$, therefore, 
including the asymptotic divergent terms $x^2$, $x$ and $\ln(x)$. The coefficients of this expansion are obtained 
analytically by equating the corresponding terms of the two sides of the equation. For the BC2 and BC23 reference 
functions, the results are reported in Eqs.~\eqref{eq:MI2} and \eqref{eq:MI23}, respectively. In the 
expansions~\eqref{eq:MI2} and \eqref{eq:MI23}, all coefficients are determined, except the integration constant 
$\mathcal M_0$ of the Ricatti equation~\eqref{eq:solLK-M}, which may depend on the reference pair. The other expansion 
coefficients depend on the multipolar coefficients $c_p$ of the long-range potential $V(x)$, on $c_{3f}$ defined in the 
single-term potential $V_f(x)$, and also on the integration constant $\mathcal M_0$. 
 
Using the ${\mathcal M}(x)$ expansions up to the order $1/x^7$, we integrate the first-order differential equation~\eqref{eq:solLK-A} for the logarithmic derivative of $\mathcal A(x)$. 
The integration constant $\mathcal A_0$ is determined by imposing  the condition $u(x)\rightarrow x^2$ for $x\rightarrow
 \infty$ upon the threshold wave function, which implies ${\mathcal A}(x)\rightarrow 1$. For the BC2 and BC23 reference pairs, the expressions of $\mathcal A(x)$ are given in Eqs.~\eqref{eq:A-CI2} and~\eqref{eq:A-CI23}, respectively, and the corresponding threshold wave functions $u(x)$ in Eqs.~\eqref{eq:u-CI2} and~\eqref{eq:u-CI23}.

The parameter $\mathcal M_0$ does not depend on the asymptotic form of the potential and accounts for the interactions at short range. Since ${\mathcal M}_0$ 
is the value in reduced units of a quantity that has dimension of volume, it is a good candidate for defining a generalized scattering volume, except for its dependence on the chosen reference functions. We show next that in fact ${\mathcal M}_0^{BC2}$ and ${\mathcal M}_0^{BC23}$  provide equivalent descriptions of the short-range interactions. Equalizing the coefficients of the  $1/x^q$ and $\ln(x)/x^q$  terms 
in the expansions~\eqref{eq:u-CI2} and \eqref{eq:u-CI23} of $u(x)$ leads to the following unique relation
\begin{eqnarray}
{\mathcal M}_0^{BC23}-{\mathcal M}_0^{BC2}&=&-\frac{2}{9}c_{3f}c_4-\frac{11}{144} c_3^2c_{3f}-\frac{1}{24}c_3c_{3f}^2\, \nonumber \\
&&+\left(\frac{83}{432}-\frac{\gamma}{6}-\frac{\ln(c_{3f})}{12}\right)c_{3f}^3\,, 
\label{eq:M0CI2-M0CI23}
\end{eqnarray}
where $\gamma$ denotes the Euler constant. 
Note that we have verified the uniqueness of this relation~\eqref{eq:M0CI2-M0CI23} for terms up to $q=5$.
The difference $\Delta \mathcal M_0=\mathcal M_0^{BC23}-\mathcal M_0^{BC2}$ depends on the parameters $c_3$, 
$c_4$,  and  $c_{3f}$, and is perfectly known as soon as the reference pairs are chosen. Therefore, it is sufficient to 
determine $\mathcal M_0^{BC2}$, because $\mathcal M_0^{BC23}$ is known unambiguously once  $c_{3f}$ is fixed. In 
particular $\mathcal M_0^{BC23}$ and $\mathcal M_0^{BC2}$ diverge simultaneously. The divergences, the most 
important features in scattering, indicate  a quasi-resonant situation with a bound state located just at threshold.
They are associated to infinite contact interactions in a pseudopotential technique describing the short-range interaction of 
the two particles by $\ell$-wave contact potentials~\cite{Derevianko05,Idziaszek06}.

So far, the analytic solutions are derived in a single-channel approximation~\eqref{eq:Schr} to the coupled channels asymptotic Schr\"odinger equation~\eqref{eq:asy}. The dipole-dipole interaction in~\autoref{eq:asy} directly couples the $\ell$ and $\ell\pm 2$ channels.
As a consequence, the contribution of  the $\ell=1+2q$ channel to the $p$-wave effective 
potential~\eqref{eq:pot} appears at order $q$ of perturbation theory and increases 
as $\mathcal I^q/x^{q+2}$.

For the considered range of non-resonant intensities $\mathcal I < 40 $ ru, a 2-channel treatment including $\ell=1$ and $3$ provides a good approximation of the effective field-dressed $p$-wave  potential. We use three different effective potentials approximating the multipolar expansion~\eqref{eq:pot}, which are given in~\autoref{tab:pot} of  appendix~\ref{app:subsec-adiab-pot}. The simplest one is the diabatic potential $V^m_d(x)$, which
is equal to the $\ell=1$ diagonal matrix element. The second one is the effective adiabatic potential $V^m_{ad}(x)$ obtained as the lowest eigenvalue of the $2\times 2$ potential matrix accounting for the coupling between the $\ell=1$ and $\ell=3$ channels. The third one is the $V^m_{n ad}(x)$ potential, which adds to $V^m_{ad}(x)$ the diagonal contribution of the non-adiabatic coupling that arises from the $x$-dependence of the adiabatic eigenvector $\Psi(x)$, the so-called 'kinetic energy' term $\langle\Psi\,|\,d^2\Psi/dx^2\rangle$. For the considered intensities, $V^m_{nad}(x)$ with multipolar coefficients up to $q=6$ describes rather well the effective potential in the $p$-wave at distances $x>20$ ru. These approximations are used in~\autoref{subsec:fits} to  numerically evaluate the coefficients of the analytical single-channel Levy-Keller formula and to analyze the asymptotic behavior of $\mathcal M(x)$ calculated in multi-channel models. This allows us to justify the procedure developed below to extract from 
$\mathcal M(x)$ a constant term similar to $\mathcal M_0$, 
characterizing the interactions at short range.
%
\section{Determining the generalized scattering volume from multi-channel asymptotic calculations}
\label{sec:numerical}
%
In this section, we determine the field-dressed generalized scattering volume $\mathcal{M}_0$ in the multi-channel case. By using an asymptotic model and the nodal line technique~\cite{CrubellierPRA17} %
(reviewed briefly in Appendix~\ref{app:subsec-M0}), 
 we calculate numerically, in Sec.~\ref{subsec:fits}, the tangent of the local phase $\mathcal M(x)$ from the $p$-wave of 
 the threshold wave function, diverging only in this channel. Then, we expand $\mathcal M(x)$ in an analytical form 
 involving the $x$-dependent terms suggested by the single-channel Levy-Keller approach. A numerical fitting 
 procedure (explained in Appendix~\ref{app:subsec-fits}) allows to extract from $\mathcal M(x)$ a constant similar to $\mathcal{M}_0$ occurring in the single-channel model. This constant depends only on the short-range interactions, represented in the model through the so-called nodal parameter $x_{00}$,
which is the position of one of the most outer nodes of the field-free $s$-wave function at threshold.
This single parameter 
determines the position of the $\ell$-, energy- and field-intensity-dependent 
nodal lines used in a universal model~\cite{LondonoPRA10,CrubellierNJP15a}. 
The constant  $\mathcal{M}_0$ is identified as the generalized field-dressed scattering volume in~\autoref{subsec:fits}. 
A detailed comparison of the numerical multi-channel results with the analytical single-channel Levy-Keller ones justifies the procedure used to calculate this generalized scattering volume. 
Its dependence on the field-free scattering length
or equivalently on the nodal parameter $x_{00}$ displays 
a divergence each time a bound state is located at threshold (see~\autoref{subsec:long-vol}). A similar resonance structure occurs when the non-resonant light intensity varies. 
%
\subsection{Asymptotic behavior of ${\mathcal{M}}(x_{max})$ from numerical calculations}
\label{subsec:fits}
%
By using the nodal line technique, the asymptotic multi-channel Schr\"{o}dinger equation~\eqref{eq:asy} is solved at threshold ($k=0$) for given magnetic quantum number $|m|$ and light intensity $\mathcal I$, or equivalently a strength $\mathcal D$ of the dipole-dipole interaction 
defined in Eq.~\eqref{eq:corresp-D}. The nodal line technique is described in Appendix~\ref{app:subsec-M0} and  Ref.~\cite{CrubellierPRA17}. 
One imposes upon the searched particular  wave function the condition to vanish asymptotically in all $\ell>1$ channels 
and to be a combination of $x^2$ and $1/x$ functions in the $p$-channel. From the numerical $p$-wave component of this 
solution,
we extract the relative weight ${\mathcal{M}}(x)$ of the asymptotic $x^2$ and $1/x$ contributions, which depends on the 
nodal parameter $x_{00}$, \ie, on the short range properties, and on the starting point of the inward integration $x_{max}$. 
Fixing the nodal parameter $x_{00}$, the asymptotic behavior of ${\mathcal{M}}(x_{max})$ is analyzed by varying the 
starting point $x_{max}$ of the inward integration. This analysis is done using the analytical single-channel Levy-Keller 
expressions of $\mathcal M(x)$, cf. Eqs.~\eqref{eq:MI2} and~\eqref{eq:MI23}. Since the effective $p$-wave potential 
decreases as $-|c_3]/x^3$, $\mathcal M(x)$ diverges as $x^2$ with increasing $x$. According to 
the expansion of the analytical function $\mathcal M(x)$, we expand the numerical function $\mathcal M(x_{max})$ in 
terms of $1/(x_{max})^q$ and $\ln x_{max}/(x_{max})^q$. For BC2, we consider the following terms (written in $x$ instead 
of $x_{max}$)
\begin{subequations}
\label{eq:fit} 
\begin{align}
&x^2,\,x,\,\ln(x),\,1,\,\frac{\ln(x)}{x},\,\frac{1}{x},\,\frac{\ln(x)}{x^2},\,\frac{1}{x^2},\dots\,.
\label{eq:fitCI2}
\intertext{For BC23 and $|m|=1$, we have the same terms~\eqref{eq:fitCI2} as for BC2, whereas 
 for BC23 and $m=0$, we have the following ones}
&
x,\,\ln(x),\,1,\,\frac{1}{x},\,\frac{\ln(x)}{x^2},\,\frac{1}{x^2},\ldots\,, 
\label{eq:fitCI230}
\end{align}
\end{subequations}
For given $x_{00}$ and light intensity, we perform numerical fits in $x_{max}$ using combinations of the terms above to determine the corresponding coefficients in the expansion of
$\mathcal M(x_{max})$. The $x_{00}$ and intensity-dependencies are determined
by performing new multi-channel calculations and new numerical fits for any new light intensity or $x_{00}$.

For $m=0$ and $\pm 1$, $\mathcal M(x_{max})$ has been computed using $n=3$ channels with $\ell=1$, $3$, $5$, three intensities 
$\mathcal I$, and $0.142152$~ru $\le x_{00}\le 0.152135$~ru, 
the field-free scattering length, 
quasi-periodic function of $x_{00}$,   varying from 
$-\infty$ to $+\infty$ in this interval. The inward integration has been initialized with the  BC2 and BC23 boundary 
conditions (see Table~\ref{tab:fct}), and $x_{max}$ is varied in the range $[20,500]$~ru. 
The numerical coefficients of the ${\mathcal M}(x_{max})$ fits are reported in~\autoref{tab:fits} of 
appendix~\ref{app:subsec-fits}. For fixed $m$, $\mathcal I$ and BC conditions, we find
some coefficients to be independent of $x_{00}$, 
whereas others depend on $x_{00}$. 
The numerically obtained $x_{00}$-independent coefficients have been compared to the analytical ones deduced from the 
single-channel Levy-Keller approach using the multipolar coefficients $c_p$ of the adiabatic $p$-wave 
potential $V^m_{ad}(x)$, given in~\autoref{tab:pot} of Appendix~\ref{app:subsec-adiab-pot} and 
the $c_{3f}$ coefficient for the BC23 boundary condition. 
We find  sgood agreement between the numerical and analytical results. 
The  $x_{00}$-dependent coefficients, 
labeled as $ {v}_m(\mathcal I,x_{00})$ and $ {\eta}_m(\mathcal I,x_{00})$,
are the prefactors of the constant and $1/x$ terms, respectively. They display the same characteristic $x_{00}$-dependence with several divergences. In fact, as expected from the Levy-Keller approach, (see Eqs.~\eqref{eq:MI2} and~\eqref{eq:MI23}),  these coefficients are related as $ {\eta}_m(\mathcal I,x_{00})={\mathrm \alpha} \times {v}_m(\mathcal I,x_{00}) +{\mathrm \beta}$, with ${\mathrm \alpha}$ and ${\mathrm \beta}$ not depending on $x_{00}$ 
because they can be expressed analytically
 in terms of the $c_p$ and $c_{3f}$ coefficients. Here also these  $x_{00}$-independent 
 numerically obtained coefficients, ${\mathrm \alpha}$ and ${\mathrm \beta}$,   agree with the analytical ones.   
For given $m$ and $\mathcal I$, the curves $ {v}_m(\mathcal I,x_{00})$ associated to the constant term 
of the expansion and obtained with the BC2 and BC23 reference functions differ by a quantity that does not depend on 
$x_{00}$.
The numerical value of this difference  is in good agreement with the
 value of the analytical formula 
$\Delta \mathcal M^0=\mathcal M^0_{BC23}-\mathcal M^0_{BC23}$~\autoref{eq:M0CI2-M0CI23} 
calculated using  the $c_p$ and $c_{3f}$ coefficients. 

We can interpret, in the multichannel numerical calculations, the coefficients of the  $\mathcal M(x_{max})$  asymptotic expansion in a satisfactory way by comparison with the $\mathcal M(x)$ expansion in the analytical single-channel Levy-Keller approach in the following way.
In detail, we use  the coefficients of the adiabatic $p$-wave potential (see Appendix~\ref{app:subsec-adiab-pot}) and the 
asymptotic boundary conditions deduced from the potential $V_f(x)$ 
 in the Levy-Keller approach (see Appendix~\ref{app:subsec-LK-eq}).   This justifies the identification of the constant term ${v}_m(\mathcal I,x_{00})$ of this multichannel calculations with the field-dressed generalized scattering volume. By analogy to the definition deduced from the single-channel Levy-Keller approach, the label $\mathcal M_0(\mathcal I,x_{00})$ is introduced.

This fairly good agreement between analytical and numerical calculations proves that, for the considered light intensities, the adiabatic potential $V^m_{nad}(x)$ represents well the effective potential in the $p$-wave. More importantly, it corroborates the separation of the expansion terms of $\mathcal M(x)$ into two types, see Eqs.~\eqref{eq:MI2} and~\eqref{eq:MI23}. The first type depends only on the asymptotic potential through the multipolar coefficients $c_p$ and the coefficient $c_{3f}$ for BC23 conditions. Among the second type, we encounter the constant coefficient 
${v}_m(\mathcal I,x_{00})$, which depends on $x_{00}$ and is similar to the parameter $\mathcal M_0$ of the analytical approach.
In fact, the constants ${v}_m(\mathcal I,x_{00})$ obtained with the BC23 and BC2 boundary conditions are equivalent and differ simply by the known quantity $\Delta \mathcal M^0$,~\autoref{eq:M0CI2-M0CI23}. In addition, when the light intensity tends to zero, they both approach the regular field-free scattering volume.

Let us emphasize that the divergent contributions to $\mathcal M(x)$, 
which are proportional to $x^2$, $x$ and $\ln(x)$,  
arise from interactions that asymptotically decrease as $1/x^3$, $1/x^4$ and $1/x^5$. In contrast, the van der Waals interaction occurs (to leading order) in the term proportional to $1/x$ in the expansion of $\mathcal M(x)$.
Furthermore, for $\mathcal M_0$, a direct comparison between the numerical and analytical results is not possible. In fact, in the analytical Levy-Keller model, $\mathcal M_0$ appears as integration constant in the solution of~\autoref{eq:solLK-M}  (the parameter $ t_0$ in~\autoref{eq:phase-infty-a}). It could, in principle,  be obtained by imposing boundary conditions at small $x_{00}$. However, since there is no explicit expression for this boundary condition, it does not provide a means to determine $\mathcal M_0$. Therefore, the $x_{00}$-dependence can be determined  only from numerical calculations  in which all the interactions present at $x<x_{00}$ are captured by the nodal parameter whereas the asymptotic Hamiltonian~\eqref{eq:asy} explicitly accounts for the interactions occurring at $x>x_{00}$. In the numerical approach, $\mathcal M_0$ is perfectly determined provided that the number of channels $n$ is sufficiently large to ensure convergence, as will be shown in~\autoref{subsec:long-vol}  below.

\subsection{Dependence  of the generalized scattering volume on  field-free $s$-wave scattering length and light intensity}
\label{subsec:long-vol}
%
%
\begin{figure}[tb]
  \centering 
	\includegraphics[width=0.99\linewidth]{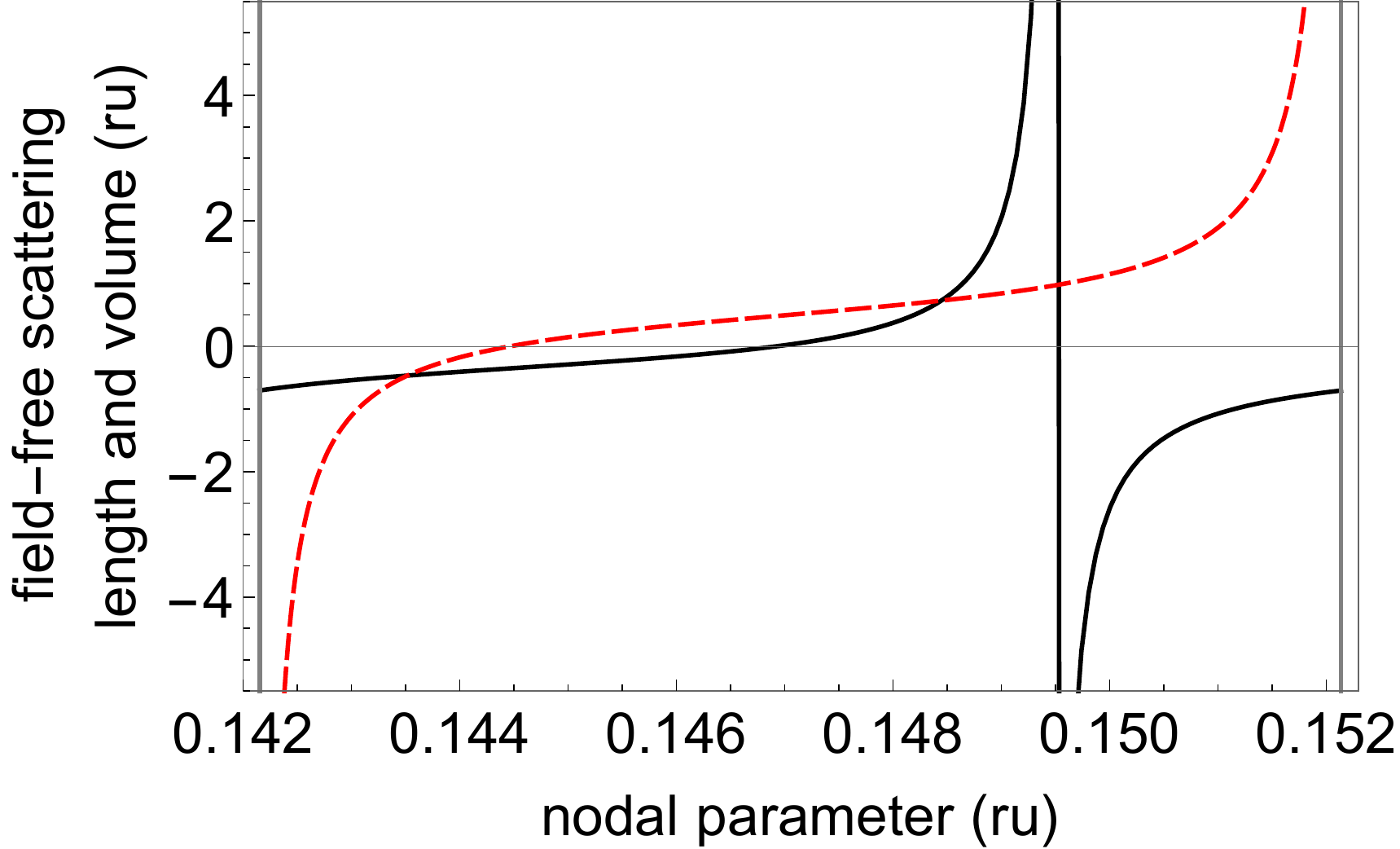}
  \caption{Field-free $s$-wave scattering length (red dashed line) and  field-free $p$-wave scattering volume 
  (black solid line) as a function of the nodal parameter $x_{00}$. The  grey vertical lines indicate the limits of the quasi-period of the variation of the field-free $s$-wave scattering length associated with the seventh node (counted from outside) of the field-free $s$-wave threshold wave function. A variation of the nodal parameter corresponds to e.g. a change of collision partners.}
  \label{fig:scatt-free}
\end{figure}
%
%
\begin{figure}[tb]    
\centering
	\includegraphics[width=0.99\linewidth]{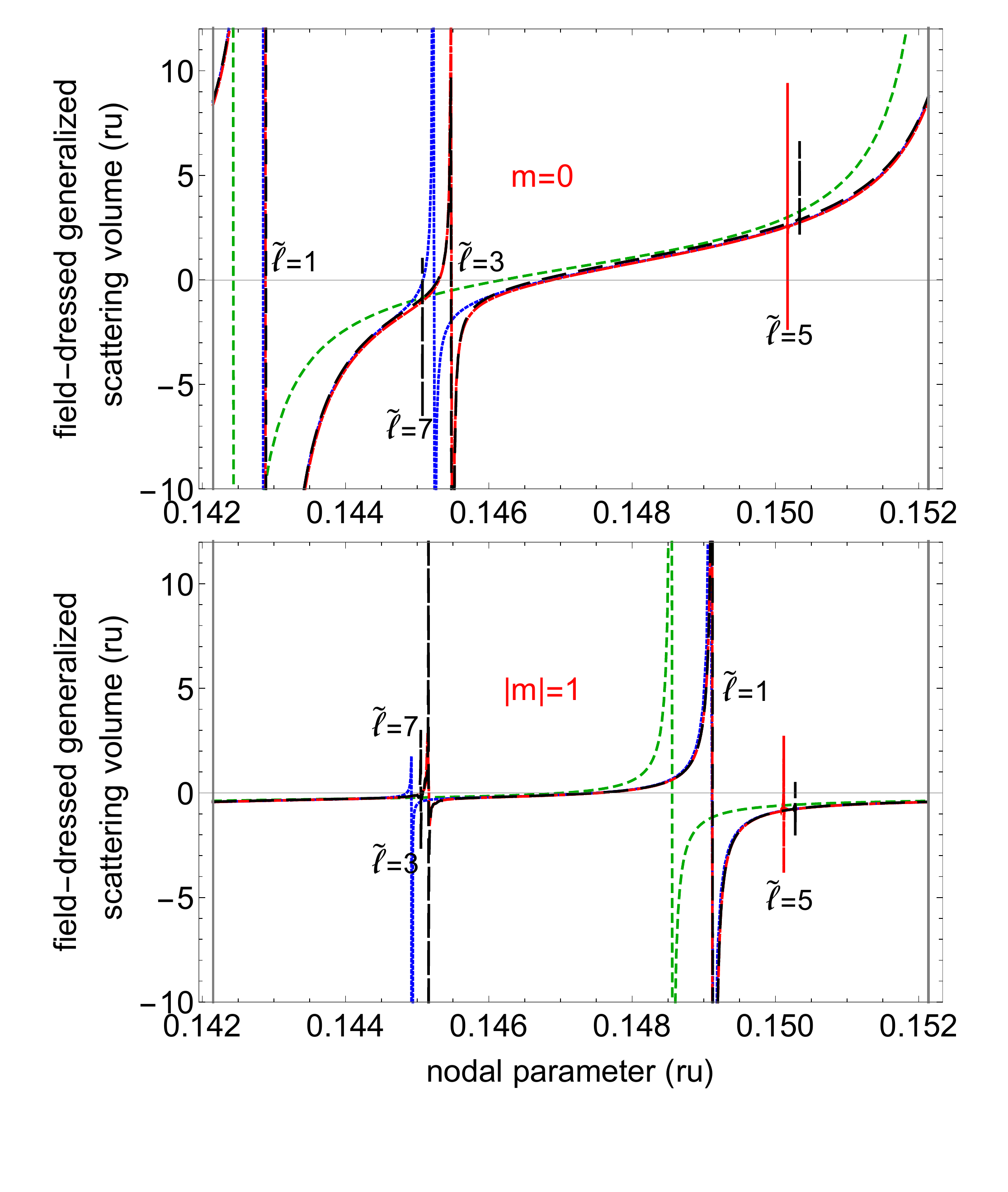}
   \caption{Field-dressed generalized $p$-wave scattering volume for $m=0$ (upper pannel) and $|m|=1$ (lower pannel) 
   as a function  of the nodal parameter $x_{00}$ calculated for $n=1$ (green dashed line), $n=2$ (blue dotted line), $n=3$ (red dot-dashed line) and $n=4$ (black solid line) channels.
  The non-resonant laser intensity is fixed at  $\mathcal I=6$~ru. 
     The resonances labeled by $\widetilde\ell$ correspond to a bound state at threshold with a dominant $\ell$-contribution in its wave function.
     The vertical gray lines correspond to an infinite $s$-wave scattering length.}
  \label{fig:scatt-dim}
\end{figure}
%
For a given light intensity, the  generalized $p$-wave scattering volume depends only on the field-free $s$-wave scattering length, or on the nodal parameter $x_{00}$,
and captures all effects of the short-range interactions. 
We recall here that, as soon as the $s$-wave scattering length of a colliding atomic pair is known, we can fix a suitable $x_{00}$ for modeling the system using the 
asymptotic model~\cite{LondonoPRA10,CrubellierPRA17}. 
Figure~\ref{fig:scatt-free} shows the field-free $s$-wave scattering length and field-free $p$-wave scattering volume versus 
$x_{00}$. In the absence of the non-resonant light, we use the ordinary definition of the scattering 
volume~\eqref{eq:def_a_ell} because the anisotropic $1/x^3$-term in the asymptotic Schr\"odinger 
equation~\eqref{eq:asy} vanishes. Furthermore, there is no channel coupling and the single-channel approximation 
becomes exact. The range for $x_{00}$ in~\autoref{fig:scatt-free}, $x_{00}\in[0.142152,\,0.152135]$~ru, corresponds to 
one quasi-period of the $s$-wave scattering length varying from $-\infty$ to $+\infty$. The divergences of the $s$-wave 
scattering length are experimentally important, since they represent a resonant situation with an $\ell=0$ bound state at 
threshold~\cite{Messiah69,Joachain83}, and correspond to the contact interaction becoming infinite. The $p$-wave 
scattering volume also displays a singularity within this $x_{00}$ range, with an $\ell=1$ bound state at 
threshold. The position and width of the singularities of the scattering length and scattering volume are completely 
different. This has been observed before for a truncated $x^{-6}$ potential, with a repulsive wall at the position 
$x_0 \rightarrow 0_+$~\cite{GaoPRA09}, a simple model that predicts the $s$-wave scattering lengths corresponding to 
divergences of the scattering parameters~\eqref{eq:def_a_ell} in any partial $\ell$-wave. In particular, the field-free 
resonances $\ell$ and $\ell'=\ell+4q$ ($q$ integer) were found to be degenerate~\cite{GaoPRA09}. For the 
studied $x_{00}$-range, our asymptotic model predicts that the field-free $\ell=1$ scattering volume  diverges at 
$x_{00}=0.1495$, \ie, for  the field-free scattering length $a=0.9668$~ru (see also Ref.~\cite{LondonoPRA10}), instead of 
the universal value $0.96$~ru~\cite{GaoPRA09}. Similarly, the $\ell=3$ field-free scattering volume, not shown 
in~\autoref{fig:scatt-free}, diverges for $x_{00}=0.1447$~ru, \ie, for $a=0.05651$~ru, instead of the universal value 
$a=0$~ru given in Ref.~\cite{GaoPRA09}.

We now analyze the field-dressed scattering volume as a function of $x_{00}$ for $m=0$ and $m=\pm1$ 
at a weak intensity $\mathcal I=6$~ru in~\autoref{fig:scatt-dim}. Note that the non-resonant field interaction removes the $m$-degeneracy. 
For the  single-channel model $n=1$ and $m=0$, the generalized field-dressed scattering volume diverges at 
$x_{00}=0.1427$~ru in the top panel of ~\autoref{fig:scatt-dim}, which is shifted from the field-free position 
$x_{00}=0.1497$~ru 
in~\autoref{fig:scatt-free}. This large shift in $x_{00}$, $\Delta x_{00} \sim 0.007\,$ru, shows the high sensitivity of the 
resonance on the non-resonant field intensity, due to the low rotational barrier for $\ell=1$. By increasing the 
number of channels to $n=2$, the $\ell=1$, $m=0$ resonance is slightly shifted toward higher $x_{00}$ values, up to $x_{00}\sim 0.1432$~ru. This is due to a small contribution of the $\ell=3$ channel to the bound state wave function now labeled by $\widetilde \ell=1$. For $\mathcal I=6$~ru, the different channels are weakly coupled so that the bound states $\widetilde \ell$ are clearly characterized by their dominant $\ell$-wave contribution. The position of the $\widetilde \ell=1$ resonance slightly varies with increasing $n$, and is stabilized for $n\ge3$. For the $n=2$ calculation, the scattering volume shows an additional singularity at $x_{00}=0.1452$~ru associated with the occurrence of a bound state at threshold with dominant $\ell= 3$ weight. As expected, the position of the field-dressed $\widetilde\ell=3$ resonance is close to the  field-free one $x_{00}=0.1447$~ru, the small shift  $\Delta x_{00} < 0.005$ ru is due to the higher rotational barrier.
For $n=3$, another singularity associated with a $\widetilde \ell=5$ bound state occurs around $x_{00}=0.150$~ru, and is very close to the field-free $\ell=1$ resonance shown in Fig.~\ref{fig:scatt-free}. By using 4 channels, a $\widetilde \ell=7$ resonance  appears at $x_{00}\approx 0.145$~ru, very close to the $\widetilde\ell=3$ one. Indeed, by increasing the number of channels, the predicted degeneracy of the $\widetilde \ell$ and $\widetilde\ell+4$ resonances becomes manifest.

We analyze now the field-dressed resonances with $|m|=1$ shown in the bottom panel of~\autoref{fig:scatt-dim}. A comparison with the field-dressed $m=0$ resonances shows significant differences. In the $n=1$ channel model, the  $|m|=1$ resonance position is $x_{00}=0.1486$~ru, suffering a shift, $\Delta x_{00}=0.0009$~ru, smaller than the one we encounter for the  $|m|=0$ resonance. Furthermore, the field-dressed $|m|=1$ resonance is much narrower than the $m=0$ one. This is ascribed to the effective potentials being asymptotically attractive and repulsive for $m=0$ and $|m|=1$, respectively, and the latter having a $1/x^3$ contribution two times smaller (see~\autoref{tab:pot}). By increasing the number of channels, additional resonances with $|m|=1$ and $\ell \ge 3$ appear. This demonstrates that, for these resonances, the strength of the channel mixing is approximately independent of $m$. Indeed, as
$\ell$ increases, the resonance positions become almost $m$-independent, since they are essentially governed by the height and width of the rotational potential barriers.

At this rather low intensity $\mathcal I$=$6$~ru, the calculation of the field-dressed resonance $\widetilde \ell$ is almost 
converged when the multi-channel model includes 
up to the $\ell'=\ell+2$ channel, which corresponds to a model including $n \ge (\ell+3)/2$ channels. 
In contrast, a larger number of channels is needed  for much higher intensities or for dipolar partners coupled by
strong dipole-dipole interaction $\cal D$. For instance, in Ref.~\cite{BohnNJP09} more than 30 channels are needed to describe the scattering cross sections of aligned dipolar molecules at ultracold collision energies. Indeed, collisions between KRb (resp. RbCs) molecules with equivalent dipole length $\mathcal D\sim 5700~a_0$ (resp. $47000~a_0$) and van der Waals length $\sigma\sim 140$ (resp. $180~a_0$) correspond to collisions in a strong non-resonant light with very high intensity $\mathcal I=240$~ru (resp. $1600$~ru).


\section{Conclusions}
\label{sec:conclusion}
%
The $p$-wave scattering volume is known to diverge for $1/R^3$ interactions, which appear for atoms in a non-resonant laser field or for the dipole-dipole scattering between ultracold atoms or molecules. In this work, we have defined a generalized $p$-wave scattering  volume for two trapped ultracold atoms in non-resonant light. To this end, we have employed an asymptotic model~\cite{LondonoPRA10,CrubellierPRA17}, based on the fact that ultracold collisions are dominated by long-range forces. The short-range interactions are taken into account by the nodal parameter, which is fixed once the field-free $s$-wave scattering length is known.

%
In detail, the threshold $p$-wave wave function is analytically constructed in a single-channel approximation using the two-potential approach developed by Levy and Keller~\cite{LevyJMP63}. For increasing $R$, this wave function expresses as a linear combination of $R^2$ and $1/R$, the free motion solutions, and the later term multiplied by quantity similar to a phase shift, which asymptotically diverges. 
The analytical $R$-expansion of this quantity contains a single constant term depending only on the nodal parameter and characterizing the short range interactions, which is identified as the generalized scattering volume. In the multi-channel calculations, the $p$-wave scattering volume is obtained by fitting unambiguously the asymptotic behavior of the $p$-wave component to its analytical expression. 

The asymptotic model depends only on the nodal parameter, which is fixed once the field-free $s$-wave scattering length 
of the collision partners is known~\cite{LondonoPRA10}. 
For a given non-resonant light intensity, we have analyzed the dependence of the field-free $p$-wave scattering volume on 
this parameter, which is significantly different from the $s$-wave scattering length dependence. The $p$-wave scattering 
volume also displays one singularity in the parameter range where the $s$-wave scattering length changes from $-\infty$ 
to $+\infty$. This is in line with earlier predictions~\cite{GaoPRA09}. The singularity is caused by the appearance of a 
$\ell=1$ bound state at threshold. In the non-resonant light, the original $p$-wave singularity is shifted and, more 
remarkably, additional singularities appear. This is due to the field-dressed $p$-wave containing contributions from 
additional field-free partial waves for which a bound state at threshold appears.
%

Instead of universal nodal lines with a single nodal parameter introduced in this paper it is possible to consider nodal lines with energy-, $\ell$- and also intensity- (equivalently dipole strength)-dependence adjusted to a real pair of atoms~\cite{LondonoPRA10,CrubellierNJP15a}. In this case, the short-range interactions are more precisely accounted and an accurate prediction of the near threshold resonances becomes possible. This description is equivalent to those introducing a regularized zero-range potential, the so called contact interaction, with infinitely many terms~\cite{Derevianko05}, but is  probably more tractable. In addition, ultracold collisions ($k\ge 0$) can be studied in a straightforward extension of this work to the non-zero energy regime, as it has been shown in previous studies devoted to the analysis of shape resonances~\cite{LondonoPRA10,CrubellierNJP15a,CrubellierNJP15b}.

In the following paper (Paper II)~\cite{Crubellier18b}, we  will use the method developed here to control the generalized scattering volume using non-resonant light. This is an extension of our previous work on controlling the s-wave scattering length~\cite{CrubellierPRA17}.
%
%
%
\begin{acknowledgments}
Laboratoire Aim\'{e} Cotton is   
"Unit\'e mixte UMR 9188 du CNRS, de l'Universit\'e Paris-Sud, de 
l'Universit\'e Paris-Saclay et de l'ENS Cachan", member of the
"F\'{e}d\'{e}ration Lumi\`{e}re Mati\`{e}re" (LUMAT, FR2764) and of
the "Institut Francilien de Recherche sur les Atomes Froids" (IFRAF).
R.G.F. gratefully acknowledges  financial support by the Spanish
 Project No.  FIS2014-54497-P (MINECO), and by the Andalusian research
group FQM-207. 
\end{acknowledgments}
%
%
\appendix
%
\section{Asymptotic Schr\"{o}dinger equation in reduced units}
\label{app:red-units}
%
Within the Born-Oppenheimer approximation, the asymptotic Hamiltonian  describing the relative nuclear motion of
 two atoms in a non-resonant light polarized along the laboratory $Z$-axis reads~\cite{CrubellierPRA17}
\begin{equation}
  \label{eq:2D_Hamil}
  H =   T_R+\frac{\hbar^2\mathbf{L}^2}{2\mu  R^2} +V_g(R)
    + {\cal D}  \frac{3\cos^2\theta -1}{R^3} \,,
\end{equation}
where $R$ is the interparticle distance, $\theta$ the angle between $\vec R$ and the  $Z$ axis, 
$\mu$  the reduced mass, $T_R$ the radial kinetic energy, and $\mathbf{L}$ the orbital angular momentum operator.
The  potential describing the short-range interaction, $V_g(R)$, is limited here to the van der Waals potential, $V_g(R)$=$-C_6/R^6$, with $C_6$ the van der Waals coefficient.
The last term in the Hamiltonian~\eqref{eq:2D_Hamil} stands for the anisotropic 
interaction 
due to the coupling of the  linearly polarized  non-resonant light with intensity $I$ and  the polarizability anisotropy 
of the particles with strength
\begin{equation}
  \label{eq:corresp-D}
 {\cal D}\,=\,
\frac{4 \pi I}{c}\alpha_1\alpha_2\,,
\end{equation}
with $\alpha_{1,2}$ being the static polarizabilities of the atoms.
This interaction is  of the dipole-dipole type, and is the same for dipoles aligned along the laboratory $Z$-axis~\cite{Crubellier18b}.

%

%

The  Hamiltonian~\eqref{eq:2D_Hamil} commutes with parity and with $L_Z$, the projection of the orbital angular momentum on the laboratory $Z$ axis.
As a result, the magnetic quantum number $m$ is conserved.  

A universal form of the Hamiltonian~\eqref{eq:2D_Hamil} is obtained by introducing reduced units. 
Here, we use the 'van der Waals reduced units' (denoted as ru) of length $x$, energy ${\mathcal E}$, and non-resonant field intensity ${\mathcal I}$ respectively defined by $R$=$\sigma x$, $E - E_0 $=$\epsilon \,{\mathcal E}$ ($E_0$ denotes the lowering of the dissociation limit), and  $I$=$\beta~{\mathcal I}$~\cite{LondonoPRA10,CrubellierNJP15a}. 
The characteristic length $\sigma$, energy $\epsilon$ and field intensity $\beta$ are
\begin{subequations}
\label{eq:scaling}
\begin{eqnarray}
\sigma & = & \left(\frac{2\mu C_6}{\hbar^2}\right)^{1/4}\,, 
\label{eq:sigma}\\ 
\epsilon & = & \frac{\hbar^2}{2\mu\sigma^2}\,,
\label{eq:epsilon} \\ 
\beta & = & \frac{c}{12\pi} \frac{\hbar^{3/2}C_6^{1/4}}{\alpha_1\alpha_2(2\mu)^{3/4}} 
=\frac{c\sigma^3\epsilon}{12\pi\alpha_1\alpha_2}\,.
\label{eq:beta}
\end{eqnarray}
\end{subequations}
%
These unit conversion factors contain all the information specific to the particle species ($\mu$,  $C_6$, 
$\alpha_{1}$ and $\alpha_{2}$). 
For a dipole-dipole interaction characterized by the strength $\mathcal D$~\eqref{eq:corresp-D}, the reduced intensity is 
$\mathcal I$=$ 3\mathcal D /\epsilon \sigma^3$. 
With these reduced units, the asymptotic Schr\"odinger equation associated to Hamiltonian~\eqref{eq:2D_Hamil} 
takes the form of~\autoref{eq:asy}. 
%
\section{Two-potential Levy-Keller approach}
\label{app:sec-Levy-Keller}
%
%
\begin{table*}[tb]
  \begin{tabular}{|l|c|c|c|c|}
    \hline
label &$k$  &pair of functions                 &  asymptotic limits & $W$               \\ \hline \hline
BC2$k$&$>0$&$kx\,j_\ell(kx)$                  &  $\sin(kx-\ell \pi/2)$ & $-k$             \\
		  &     &$-kx\,\eta_\ell(kx)$              &   -$\cos(kx-\ell \pi/2)$&                \\  \hline \hline
BC2   &$0$  &$x^{\ell+1}$                      & $x^{\ell+1}$& $-(2\ell+1)$       \\
	    &     &$x^{-\ell}$                       &       $x^{-\ell}$  &     \\ \hline \hline
BC23  &$0$  &$-\pi\,\,(c_{3f})^{\ell+1/2}/(2\ell)!\,\sqrt{x}\,Y_{2\ell+1}(\sqrt{4c_{3f}/x})$ & $x^{\ell+1}$&$-(2\ell+1)$  \\  
	    &     &$\,\,(c_{3f})^{-\ell-1/2}\,\,(2\ell+1)!\sqrt{x} \,J_{2\ell+1}(\sqrt{4c_{3f}/x})$& $x^{-\ell}$  &      \\       
\hline  
  \end{tabular}
  \caption{\label{tab:fct}  Analytical expressions of the reference pairs labeled in column 1. 
Van der Waals reduced units  ru defined in Sec.~\ref{app:red-units} are used. The linearly independent functions ($\varphi(x)$, $\psi(x)$), which are $\ell$-wave solutions of the Schr\"{o}dinger equation of the potential $V_f(x)$  at energy $\epsilon=k^2$ ($k$ specified in column 2), are presented in the first and second lines of column 3, and their Wronskian $W$  in column 5. The  BC2$k$  functions are free waves ($V_f(x)=0$) at positive energy $k>0$. The BC2 and BC23 functions are solutions at threshold $k=0$ of $V_f(x)=0$ and $V_f(x)=-c_{3f}/x^3$ with $c_{3f}>0$, respectively. The  asymptotic limits of these functions are given in the fourth column, note that the free spherical waves at threshold (BC2)  are equal everywhere to their asymptotic form. The analytical functions BC23 are given in Ref.\cite{MoritzPRA01}.}  
\end{table*}
%
\subsection{Levy-Keller method}
\label{app:subsec-LK-eq}
%
The two-potential method was proposed by Levy and Keller~\cite{LevyJMP63,OMalleyPR64,HinckelmannPRA71} to determine the single-channel wave function $u(x)\equiv u_\ell(x)$,  which is the $\ell$-wave solution of the Schr\"odinger equation~\eqref{eq:Schr}  associated to the 
potential $V(x)\equiv V_\ell(x)=-c_3/x^3-c_4/x^4-c_5/x^5 -c_6/x^6$, 
see~\autoref{eq:pot}, with energy 
$\epsilon=k^2$. In this model a second potential $V_f(x)$ is introduced, leading to the definition of a 'reference' pair of functions ($\varphi(x)$, $\psi(x)$). They are linearly independent solutions of the Schr\"odinger equation of the potential $V_f(x)$ at the same energy $\epsilon$ and for the same $\ell$. This $\ell$-wave solution $u(x)$ is written as a linear $x$-dependent combination of the reference functions 
\begin{subequations}
\label{eq:systLK}
\begin{align}
u(x)&={\mathcal A}(x)(\varphi(x)-\psi(x)\,{\mathcal M}(x))\, ,
\label{eq:fctLKap}
\\
\intertext{with the imposed condition}
\frac{du(x)}{dx}&={\mathcal A}(x)\left(\frac{d\varphi(x)}{dx}-\frac{d\psi(x)}{dx}\,{\mathcal M}(x)\right)\,.
\label{eq:fderLK}
\end{align}
\end{subequations}
In the expression~\eqref{eq:fctLKap}, $\mathcal A(x)$ is a global amplitude, and $\mathcal M(x)$ the relative amplitude of  $\varphi(x)$ and $\psi(x)$. $\mathcal M(x)$ plays the role of $\tan \delta(x)$, with $\delta(x)$ being the local phase shift describing the collisional partial waves in terms of spherical Bessel and Neumann functions. 

To solve~\autoref{eq:systLK}, we first eliminate the global amplitude $\mathcal A(x)$ in~\autoref{eq:fctLKap} by using the radial Schr\"{o}dinger equations satisfied by $u(x)$ and by the pair ($\varphi(x)$, $\psi(x)$), and the imposed condition~\autoref{eq:fderLK}.
We derive the following equation for the relative amplitude  
\begin{subequations}
\label{eq:solLK}
\begin{align}
\frac{d\,{\mathcal M}(x)}{dx}=&-\frac{V(x)-V_f(x)}{W}\big(\varphi(x)-\psi(x)\,{\mathcal M}(x)\big)^2, 
\label{eq:solLK-M}
\\
\intertext{with $W$ being the wronskien of the reference pair, $W=\varphi(x)\psi'(x) - \varphi'(x) \psi(x)$. 
The integration of this differential equation introduces a constant $\mathcal M_0$, which may depend on the reference pair.
In a second step, we obtain the differential equation for the logarithmic derivative of $\mathcal A(x)$:}
\frac{d\ln{({\mathcal A}(x))}}{dx}=&-\frac{V(x)-V_f(x)}{W}\,\psi(x)\big(\varphi(x)-\psi(x)\,{\mathcal M}(x)\big) \, , \nonumber
\\
\label{eq:solLK-A}
\end{align}
\end{subequations}
which is integrated  imposing the boundary condition $\mathcal A(x) \rightarrow 1$ for $x \rightarrow \infty$. Once $\mathcal A(x) $ and ${\mathcal{M}}(x)$ are determined, the solution $u(x)$ is obtained. We emphasize that, obviously, $u(x)$ does not depend on the choice of the second potential $V_f(x)$ nor on the reference pair. If $V(x)$ expresses as a multipolar expansion~\eqref{eq:pot}, an analytical expression for the asymptotic form of $u(x)$ can be obtained for energy $\epsilon \ge 0$ when free waves are chosen as reference functions or at threshold $\epsilon=0$ for different reference functions.

In summary, the Levy-Keller method first computes ${\mathcal{M}}(x)$, which is related to the local phase shift, and the amplitude 
$\mathcal A (x)$ is independently obtained in a second step, after introducing an arbitrary constant $\mathcal M_0$. In contrast, in the extensively used phase-amplitude method pioneered by Milne~\cite{MilnePR30}, the amplitude satisfies a nonlinear equation that is integrated first and the phase is calculated in a second step. Thus, the pair of functions amplitude and phase, which parametrizes the wave function, is not unique and does not necessarily lead to the determination of the scattering parameters \autoref{eq:def_a_ell}. Note that a direct integral representation for scattering phase shifts, based on a modified version of Milne's approach, has been recently proposed~\cite{ShuPRA18}.

Note also that the asymptotic solution of~\autoref{eq:Schr} could be constructed using perturbation theory, as done by 
Hinckelmann and Spruch with another formulation of the two-potential approach~\cite{HinckelmannPRA71}. For the long-
range part $x>d$, they consider a single multipolar potential $V(x)=-c_{pf}/x^q$, and for the short-range one $x<d$, an 
unknown potential characterized at $x=d$ by a phase $\delta_{\ell}(k,x=d)$ such as $\tan(\delta_{\ell}(k,x=d))$ increases 
as $k^{2\ell+1}$ at low energy. For $x>d$, the phase of the wave function is obtained by treating the external part of  
$V(x)$ to first order in perturbation theory, the zeroth order consisting of free spherical waves. This procedure is analogous 
to the Levy-Keller approach with the reference pair BC2$k$ and $V_f(x)=0$, see~\autoref{tab:fct}, and determining 
$\mathcal M(x)$ by first order perturbation theory. 
%
\subsection{The  reference pairs} 
\label{app:subsec-basis-function}
%
The analytical pairs $(\varphi(x), \psi(x))$ used in this work to obtain analytical solutions $u(x)$ of the Schr\"{o}dinger equation~\eqref{eq:Schr} by the Levy-Keller method are presented in~\autoref{tab:fct}. These reference functions depend on the energy and on the chosen potential $V_f(x)$. They are labeled according to the imposed asymptotic behavior, \ie,  the boundary conditions (BC). Whereas the wave function $u(x)$ does not depend on the reference pair, the relative amplitude $\mathcal M(x)$ and the global amplitude $\mathcal A(x)$ depend a priori on the chosen $\varphi(x)$ and $\psi(x)$. 

For positive energy $\epsilon=k^2>0$, we use the reference pair labeled by BC2$k$, which  corresponds to the spherical Bessel and Neumann functions describing free spherical waves. For vanishingly small wave number $k$ and not too large distance, $x\ll 1/k$, such that $kx \rightarrow 0$, the reference functions behave as $\varphi(x)\propto (kx)^{\ell+1}$ and $\psi(x)\propto (kx)^{-\ell}$.

Considering the solutions at threshold, \ie, $k=0$, the reference pair BC2 corresponds to the partial waves for free motion, \ie, $V_f(x)=0$, with functions $\varphi_{\mathrm BC2}(x)=x^{\ell+1}$ and $\psi_{\mathrm BC2}(x)=1/x^\ell$. The $p$-wave pair of BC23 functions ($\varphi_{BC23}(x)$, $\psi_{BC23}(x))$ corresponds to the solutions at threshold of the potential $V_f(x)=-c_{3f}/x^3$ with $c_{3f}>0$. These analytical functions are proportional to the Bessel functions of second kind $Y_3(\sqrt{4c_{3f}/x})$ and first kind $J_3(\sqrt{4c_{3f}/x})$, see~\autoref{tab:fct}, and  have been used for $\ell \ge 3$ in Ref.~\cite{MoritzPRA01}

For these three sets of reference functions and $\ell=1$, $\mathcal M(x)$ is the value in reduced units of a quantity that 
has dimension of volume.
%
\subsection{Analytic expansion at threshold for $p$-waves of ${\mathcal M}(x)$, ${\mathcal A}(x)$ and $u(x)$ }
\label{app:subsec-M-expand}
%
For the potential $V(x)$~\eqref{eq:pot} and a chosen pair of reference functions, see~\autoref{tab:fct}, the asymptotic expansion of $\mathcal M(x)$, solution of~\autoref{eq:solLK-M}, is obtained analytically by identifying the coefficients of the $x^q$ and $\ln(x)/x^q$ terms, see Sec.~\ref{subsec:kx-eq-0} for more details. This method does not allow the determination of the constant term 
$\mathcal M_0$, which, therefore, does not depend on the asymptotic properties of the Hamiltonian~\eqref{eq:Schr}. When the nodal line technique is used, $\mathcal M_0$ depends only on the nodal parameter $x_{00}$.

Using the BC2 reference pair ($x^2$, $1/x$), one obtains 
\begin{widetext}
\begin{eqnarray}
\label{eq:MI2}
{\mathcal M}_{\mathrm BC2}(x)= &-&\frac{c_3}{6}x^2-\left(\frac{c_3^2}{9}+\frac{c_4}{3}\right)x 
-\left(\frac{c_3^3}{12}+\frac{c_3c_4}{3}+\frac{c_5}{3}\right)\ln(x) + {\mathcal M}_0^{\mathrm BC2}+
\left(\frac{c_3^4}{18}+\frac{2c_3^2c_4}{9}+\frac{2c_3c_5}{9}\right)\frac{\ln(x)}{x}   
\nonumber
  \\
&+&\left(\frac{11c_3^4}{162}+\frac{37c_3^2c_4}{108}+\frac{2c_4^2}{9}+\frac{c_3c_5}{3}+\frac{c_6}{3}
-\frac{2c_3{\mathcal M}_0^{\mathrm BC2}}{3}\right)\frac{1}{x} +\cdots
\end{eqnarray}
%
For the BC23 reference functions associated with the potential $V_f=-c_{3f}/x^3$ ($c_{3f}>0$), it yields
\begin{eqnarray}
\label{eq:MI23}
{\mathcal M}_{\mathrm BC23}(x)=&-&\frac{c_3-c_{3f}}{6}x^2 
-\left(\frac{(c_3-c_{3f})^2}{9}+\frac{(c_{3}-c_{3f})c_{3f}}{3}+\frac{c_4}{3}\right)x \, \\
&-&\left(\frac{c_{3}^3-c_f^3}{12}+\frac{c_{3}c_4}{3}+\frac{c_5}{3}\right)\ln(x)
+{\mathcal M}_0^{\mathrm BC23} \,
+(c_3-c_{3f})\left(\frac{c_{3}^3}{18}+\frac{2c_{3}c_{4}}{9}+\frac{2c_{5}}{9}\right)\frac{\ln(x)}{x}        \nonumber \\
&+&\Bigg((c_3-c_{3f})\left(\frac{11c_{3}^3}{162}+\frac{c_3^2c_{3f}}{72}
+\frac{c_3c_{3f}^2}{135}+\frac{491c_{3f}^3}{3240}+\frac{5c_3c_4}{54}-\frac{7c_{3f}c_4}{108}\right)
\nonumber \\
 &+&\frac{2c_4^2}{9}+\frac{c_3c_5}{3}+\frac{c_6}{3} 
-\frac{(c_3-c_{3f})c_{3f}^3\gamma}{9} -\frac{(c_3-c_{3f})c_{3f}^3\ln(c_{3f})}{18}
-\frac{2(c_3-c_{3f})}{3}{\mathcal M}_0^{\mathrm BC23}\Bigg)\frac{1}{x} +\cdots \nonumber
\end{eqnarray}
\end{widetext}
with $\gamma$ being the Euler constant.

When $c_3$ is positive, it is possible to account entirely in the reference functions for the $-c_3/x^3$ attractive contribution 
to the potential $V(x)$ by setting $c_{3f}=c_3$. 
The expression of $\mathcal M_{BC23}(x)$ is then particularly simple because  
the terms $x^2$ and $\ln(x)/x$ disappear, and the $x$ term depends only on $c_4$. 
Furthermore,  for $c_{3f}=c_3$,  the $1/x$ term, 
does not have contributions from $\mathcal M_0^{BC23}$ and $\ln(c_3)$ nor from $c_3^4$.
This simple case $c_{3f}=c_3$ is used in the study of the $\ell=1$ and $m=0$ states for which the adiabatic approximation to the effective potential in the $p$-channel is asymptotically attractive (see~\autoref{tab:pot}). For the $\ell=1$ and $|m|=1$ states, the adiabatic $p$-wave potential $V_{nad}^{|m|=1}(x)$ is repulsive ($c_3<0$ see~\autoref{tab:pot}) and $c_{3f}=-c_3$ is used in BC23 to ensure real reference functions and real $\mathcal M(x)$. The asymptotic form for $\mathcal M_{BC23}^{c_{3f}=-c_3}(x)$ is given by~\autoref{eq:MI23}.

For a given potential $V(x)$ Eq.~\eqref{eq:pot}, the asymptotic expansions of $\mathcal M_{BC2}(x)$ and $\mathcal M_{BC23}(x)$ depend on the multipolar coefficients $c_p$ of $V(x)$ and on the $c_{3f}$ coefficient defining the BC23 reference pair. Furthermore, they introduce a priori different constant coefficients, $\mathcal M^{BC2}_0$ and $\mathcal  M^{BC23}_0$, which take into account the contribution of the inner part of the potential $V(x)$ not involved in the derivation of Eqs.~\eqref{eq:MI2} and~\eqref{eq:MI23}. This constant $\mathcal M_0$ is the $x$-independent term in $\mathcal M(x)$ and also appears in some $x$-dependent terms. For instance, the coefficient of the term $1/x$ can be expressed as $\eta={\mathrm \alpha}\,\mathcal M_0+{\mathrm \beta}$, where ${\mathrm \alpha}$ and ${\mathrm \beta}$ only depend on the multipolar coefficients $c_p$ of $V(x)$ and on $c_{3f}$. Whereas for the reference pair BC23 and $c_{3f}=c_3$, the coefficient of $1/x$ is independent on $x_{00}$ and only depends on $V(x)$. The calculation of the difference $\Delta \mathcal M_0 = \mathcal M_0^{BC23} - \mathcal M_0^{BC2}$ (whose result is given in Eq~\eqref{eq:M0CI2-M0CI23}) is presented below.

Using these analytical asymptotic expansions of $\mathcal M(x)$, we integrate~\autoref{eq:solLK-A}, and impose the asymptotic condition ${\mathcal A}(x) \rightarrow 1$ for $x\rightarrow\infty$, to obtain the  analytical expressions of $\mathcal A(x)$. For the BC2 and BC23 reference pairs, we encounter the following analytical expressions of $\mathcal A(x)$:   
\begin{widetext}
\begin{eqnarray}
{\mathcal A}_{\mathrm BC2}(x)=1+\frac{c_3}{3} \frac{1}{x}+\left(\frac{c_3^2}{12} +\frac{c_4}{6}\right)\frac{1}{x^2} 
+\left(\frac{c_3^3}{36}+\frac{c_3c_4}{9}+\frac{c_5}{9}\right) \frac{1}{x^3}+\cdots\, 
\label{eq:A-CI2}
\end{eqnarray}
\begin{eqnarray}
{\mathcal A}_{\mathrm BC23}(x)&=&1+\frac{c_3-c_{3f}}{3x} 
+\left(\frac{(c_3-c_{3f})^2}{12}+\frac{c_{3f}(c_3-c_{3f})}{24}+\frac{c_4}{6}\right)\frac{1}{x^2}\, \nonumber \\ 
&+&\Bigg(\frac{(c_3-c_{3f})^3}{36}+\frac{(c_3-c_{3f})^2c_{3f}}{24}+\frac{(c_3-c_{3f})c_{3f}^2}{60}
+\frac{c_4(c_3-c_{3f})}{9}+\frac{c_4c_{3f}}{36}+\frac{c_5}{9}\Bigg)\frac{1}{x^3}+\cdots\,  
\label{eq:A-CI23}
\end{eqnarray}
%
For $c_{3f}=c_3$, the expression for ${\mathcal A}_{\mathrm BC23}(x)$ is simplified because the $1/x$ contribution disappears, and the $1/x^2$ and $1/x^3$ terms depend only on $c_4$, and  on $c_5$ and $c_4c_3$, respectively. 

Using these analytical expressions of $\mathcal M(x)$ and  $\mathcal A(x)$, we obtain from~\autoref{eq:fctLKap} the following asymptotic expansions of the threshold $p$-wave function $u(x)$: 
\begin{eqnarray}
u_{\mathrm BC2}(x)&=&x^2+\frac{c_3}{2}x+\left(\frac{c_3^2}{4}+\frac{c_4}{2}\right) 
+\left(\frac{c_3^3}{12}+\frac{c_3c_4}{3}+\frac{c_5}{3}\right)\frac{\ln(x)}{x}
+\left(\frac{17c_3^3}{216}+\frac{c_3c_4}{4}+\frac{c_5}{9}-{\mathcal M}_0^{\mathrm BC2}\right)\frac{1}{x}\nonumber \\
&-&\left(\frac{c_3^4}{48}+\frac{c_3^2c_4}{12}+\frac{c_3c_5}{12}\right)\frac{\ln(x)}{x^2}\,
-\left(\frac{79c_3^4}{1728}+\frac{11c_3^2c_4}{48}+\frac{c_4^2}{8}+\frac{c_6}{4}+\frac{37c_3c_5}{144}-\frac{c_3{\mathcal M}_0^{\mathrm BC2}}{4}\right)\frac{1}{x^2}+ \cdots
\label{eq:u-CI2}
\end{eqnarray}
\begin{eqnarray}
u_{\mathrm BC23}(x)&=&x^2+\frac{c_3}{2}x+\left(\frac{c_3^2}{4}+\frac{c_4}{2}\right)
+\left(\frac{c_3^3}{12}+\frac{c_3c_4}{3}+\frac{c_5}{3}\right)\frac{\ln(x)}{x}\,\nonumber\\
&+&\Bigg(\frac{17c_3^3}{216}+\frac{c_3c_4}{4}+\frac{c_5}{9}-{\mathcal M}_0^{\mathrm BC23}-\left(\frac{2}{9}c_4+\frac{11}{144} c_3^2\right)c_{3f} 
-\frac{c_3c_{3f}^2}{24}+\left(\frac{83}{432}-\frac{\gamma}{6}-\frac{\ln(c_{3f})}{12}\right)c_{3f}^3\Bigg)\frac{1}{x}\,\nonumber\\
&-&\left(\frac{c_3^4}{48}+\frac{c_3^2c_4}{12}+\frac{c_3c_5}{12}\right)\frac{\ln(x)}{x^2} 
-\Bigg(\frac{79c_3^4}{1728}+\frac{11c_3^2c_4}{48}+\frac{c_4^2}{8}+\frac{c_6}{4} +\frac{37c_3c_5}{144}\,\nonumber\\
&+&\frac{c_3}{4}\left({-\mathcal M}_0^{\mathrm BC23}-\left(\frac{2}{9}c_4+\frac{11}{144} c_3^2\right)c_{3f}
-\frac{c_3c_{3f}^2}{24}+\left(\frac{83}{432}-\frac{\gamma}{6}-\frac{\ln(c_{3f})}{12}\right)c_{3f}^3\right)
\Bigg)\frac{1}{x^2} +\dots
\label{eq:u-CI23}
\end{eqnarray}
\end{widetext}
Let us recall that the wave function $u(x)$ does not depend on the chosen reference pair. Thus, comparing $u_{\mathrm BC2}(x)$ and $u_{\mathrm BC23}(x)$, \ie,~\eqref{eq:u-CI2}  and~\eqref{eq:u-CI23}, the coefficients of the $1/x$ and $1/x^2$ terms are equal only if the constants $\mathcal M_0^{\mathrm BC2}$ and $\mathcal M_0^{\mathrm BC23}$ are related by~\autoref{eq:M0CI2-M0CI23}, which only involves the multipolar constants $c_p$  of $V(x)$ and the coefficient $c_{3f}$ defining  the BC23 reference pair. We have verified that relation~\eqref{eq:M0CI2-M0CI23} insures the equality of the coefficients multiplying  $1/x^3$, $1/x^4$,$1/x^5$, $\ln(x)/x^3$, $\ln(x)/x^4$, and $\ln(x)/x^5$ in the wave functions $u_{\mathrm BC2}(x)$ and $u_{\mathrm BC23}(x)$. In particular, for $c_{3f}\rightarrow 0$, the reference pairs BC23 and BC2 become identical and $\mathcal M_0^{\mathrm BC23}\rightarrow \mathcal M_0^{\mathrm BC2}$.
\begin{table*}[tb]
\small
  \begin{tabular}{|c|c|c|c|c||c|c|c|c|}
    \hline
		&\multicolumn{4}{c||}{ $m=0$} & \multicolumn{4}{c|}{$|m|=1$}   \\
$V$ &$c_3$ &$c_4$&$c_5$&$c_6$ &$c_3$ &$c_4$&$c_5$&$c_6$  \\                      \hline \hline
$V^m_d$	   &${4\mathcal I}/{15}$&$0$& $0$ &$1$&  ${-2\mathcal I}/{15}$&$0$&$0$&$1$\\                      \hline
$V^m_{ad}$	   &${4\mathcal I}/{15}$&$6\mathcal I^2/875$&$-4\mathcal I^3/65625$&$1-86\mathcal I^4/20671875$&$-2\mathcal I/15$&$4\mathcal I^2/875$&$8\mathcal I^3/65625$&$1+8\mathcal I^4/6890625$\\                      \hline
$V^m_{nad}$	   &${4\mathcal I}/{15}$&$33\mathcal I^2/4375$&$-4\mathcal I^3/46875$&$1-3814\mathcal I^4/516796875$ &${-2\mathcal I}/{15}$&$22\mathcal I^2/4375$&$8\mathcal I^3/46875$&$1+472\mathcal I^4/172265625$\\                      \hline
  \hline
  \end{tabular}
		\normalsize
  \caption{\label{tab:pot} Effective potential, given as a multipolar expansion with terms $-c_q/x^q$ ($q \ge 2$), in the $p$-wave for a pair of atoms in a non-resonant light of intensity $\mathcal I$, calculated in a two-channel model ($\ell=1$ and $\ell=3$). The term $q=2$ (not shown) is the rotational term, with $c_2=-2$. The $c_q$ coefficients, with $3 \le q \le 6$, are reported for magnetic quantum numbers $m=0$ and $|m|=1$. $V^m_d(x)$ is the diagonal term of the interaction in~\autoref{eq:asy} (the diabatic potential). 
$V^m_{ad}(x)$ is the lowest eigenvalue of the two-channel Hamiltonian (the adiabatic potential). $V^m_{nad}(x)$ is the sum of the adiabatic potential and of the non-adiabatic coupling. Note that the $c_3/x^3$ contribution is attractive for $m=0$ and repulsive for $|m|=1$.}
\end{table*}
%
%
%
\section{Multi-channel determination of $\mathcal M_0$}
\label{app:sec-M0}
%
\subsection{Effective $p$-wave potential}
\label{app:subsec-adiab-pot}
%
%
For  $m=0$ and $|m|=1$, the effective potential for the $p$-channel is written, as multipolar expansion with coefficients 
$c_p$  ($3\leq p \leq 6$)  in~\autoref{tab:pot}, using different approximations. The diabatic potential $V^m_{d}(x)$ 
corresponds to the diagonal $\ell=1$ potential matrix element of the asymptotic potential Eq.~\eqref{eq:asy}. The adiabatic 
potential $V^m_{ad}(x)$ is equal to the lowest $x$-dependent eigenvalue of the $2\times2$ potential matrix coupling the 
$\ell=1$ and $\ell=3$ waves, and the adiabatic potential $V^m_{nad}(x)$ includes in addition  the non-adiabatic effects. 
%
\subsection{Multi-channel calculations}
\label{app:subsec-M0}
%
The nodal line technique presented in detail in Ref.~\cite{CrubellierPRA17} is used to numerically solve the asymptotic multi-channel Schr\"{o}dinger equation~\eqref{eq:asy} in a $n$-channel model (odd $\ell$-values, $\ell=1,\,3,...\,2n-1$). 
We expand the threshold wave functions 
$f(x,\theta,\phi)$ in terms of spherical harmonics and restrict the number of odd-parity partial waves to $n$, with $n=(\ell_{max}-\ell_{min}+2)/2$ and
$\ell_{min}=1\le\ell\le\ell_{max}$ with $\ell$ odd. The generalized scattering volume is determined by choosing a particular $p$-wave physical threshold solution. We impose to this solution, written as the vector ${\bf z}^{j=1}(x)$, with $n$ radial components $ z^{j=1}_{\ell}(x)$, to decrease asymptotically in all channels $\ell \ge 3$, and to diverge only in the $\ell=1$ channel.
 This solution is constructed from 
$n$ particular pairs of solutions $({\bf f}^j_{+}(x),{\bf f}^j_{-}(x))$ ($1 \le j \le n$) 
with radial components 
$(f^{j}_{+,\ell}(x),\,f^{j}_{-,\ell}(x))$ in the different $\ell$-channels, and with imposed asymptotic forms. 
Each pair is associated with a 
particular channel $\ell=\ell_j$, with $\ell_{j=1}$=$1$ and the asymptotic form of its component in this $\ell_j$-channel is imposed at 
$x_{max}$ to be one of the analytical functions BC2 or BC23  defined for $k=0$ in~\autoref{tab:fct}. In other words, 
the asymptotic form in this channel is $\varphi(x)\propto x^{\ell_j+1}$ for $f^j_{+,\ell_j}(x)$ or $\psi(x)\propto 1/x^{\ell_j}$ 
for $f^j_{-,\ell_j}(x)$, whereas the components in the other channels $f^j_{\pm,\ell}(x)$ $\ell \neq \ell_j$ are vanishingly 
small.
Thus, the asymptotic form  of this solution in the $\ell=1$ channel has  to satisfy (cf.  Eqs.~(14) and (15) of Ref.~\cite{CrubellierPRA17})
\begin{equation}
\label{eq:voldif-fct}
z^{j=1}_{\ell=1}(x) = f^{j=1}_{+,\ell=1}(x) - \sum_{j'=1}^n {\bf \overline M}^{j=1}_{j'}(x_{00},x_{max}) f^{j'}_{-,\ell=1}(x)\,.
\end{equation}
where $f^{j=1}_{+,\ell=1}(x)$ increases asymptotically as $x^2$, whereas  $f^{j'}_{-,\ell=1}(x)$ vanish at least as $1/x^3$ for $j' \ge 2$ and as $1/x$ for $j'=1$, \ie,
they satisfy either the BC2 or the BC23 boundary conditions at $x_{max}$ specified in~\autoref{tab:fct}. 
For the boundary condition BC23, the potential $V_f(x)=-|c_3|/x^3$ is used to determine the initial value for the inward interaction of the pair $f^{j=1}_{\pm,\ell=1}(x_{max})$, the adiabatic potential for $p$-wave being attractive (resp. repulsive) for $m$=$0$ (resp. $|m|$=$1$) (see Table~\ref{tab:pot}).

The coefficients ${\bf \overline M}^{j=1}_{j'}(x_{00},x_{max})$ in~\autoref{eq:voldif-fct} are determined by imposing to each radial $\ell$-component of ${\bf z}^{j=1}(x)$ to vanish at short range on what we call the nodal line $x_{00}$. 
The nodal line technique~\cite{CrubellierNJP15a}  replaces the interaction at very small distances $x<x_{00}$ by a 
repulsive wall in each channel at $x_0\equiv x_0(\mathcal E,\ell,\mathcal I)$ with $x_{00}=x_0(0,0,0)$~\cite{CrubellierEPJD99,VanhaeckeEPJD04}. 
This nodal parameter $x_{00}$ 
determines the position of $\ell$-, energy- and intensity-dependent repulsive walls $x_0({\mathcal E},\ell,{\mathcal I})$ in all channels, and thus contains in an effective way all information on the short-range interaction up to the nodal line. 
For more details on the choice of $x_0({\mathcal E},\ell,{\mathcal I})$, 
the reader is referred to Ref.~\cite{CrubellierPRA17}. 
The terms ${\bf \overline M}^{j=1}_{j'}(x_{00},x_{max})$ are $x$-independent constants, and  depend on $x_{max}$, the 
starting point of the inward integration, on the nodal parameter $x_{00}$ and on the boundary conditions BC2 or BC23. 
At $x=x_{max}$, we write
\begin{eqnarray}
z^{j=1}_{\ell=1}(x_{max})\approx f^{j=1}_{+,\ell=1}(x_{max}) - \mathcal M(x_{max})f^{j=1}_{-,\ell=1}(x_{max})\,, \nonumber
 \\
\label{eq:voldif1}
\end{eqnarray}
replacing ${\bf \overline M}^{j=1}_{j'=1}(x_{00},x_{max})$ by $\mathcal M(x_{max})$. 
If we identify
\begin{eqnarray}
f^{j=1}_{+,\ell=1}(x_{max})&=&\varphi(x_{max}), \nonumber\\
f^{j=1}_{-,\ell=1}(x_{max})&=&\psi(x_{max}), 
\label{eq:voldif2}
\end{eqnarray}
Equation~\eqref{eq:voldif1} resembles the function $u(x)/\mathcal A(x)$ of the single-channel 
approximation~\eqref{eq:fctLK}, suggesting that $\mathcal M(x_{max})$ plays a role similar to the tangent of the phase 
shift in the $p$-wave at the position $x_{max}$.

\subsection{Fits of the numerical $\mathcal M(x_{max})$ to the Levy-Keller expansions }
\label{app:subsec-fits}
%
%
\begin{table*}[tb]
\small
  \begin{tabular}{|c|c|c|c|c|c|c|c|c|c|c|}
    \hline
${\mathcal I}$&$m$&BC&$x_{max}^2$&$x_{max}$&$\ln(x_{max})$&constant&$\ln(x_{max})/x_{max}$&$1/x_{max}$&$\alpha$&$\beta$      \\   
(ru)& & &$\mathcal I$&$\mathcal I^2$&$\mathcal I^3$ & &$\mathcal I^4$& &$\mathcal I$&$\sim\mathcal I^4$                \\ \hline \hline
$6$ & 0    &BC2 &$-0.26667$     &$-0.3432$      &$-0.457$      
&${v}_0(\mathcal I,x_{00})$           &$0.50$    &${\eta}_0(\mathcal I,x_{00})$     &$-1.05$      & $0.835$       \\ 
    &      &    &${-0.26667}$&${-0.3667}$&${-0.469}$ 
&${\mathcal M}_{BC2}^0$               &${0.50}$& -                                &${-1.07}$& ${1.00}$  \\  \hline
$6$ & 0    &BC23& 0             &$-0.058775$    &$-0.115$      
&${v}_0(\mathcal I,x_{00})$-0.31      & -       & 0.55                             & -           & -             \\ 
    &      &    &${0}$       &${-0.08229}$&${-0.127}$ 
&${\mathcal M}_{BC2}^0{-0.339}$     & -      &${0.496}$                      & -           & -             \\  \hline
$6$ &$\pm1$&BC2 &$0.13333$      &$-0.110295$  & $0.07625$  
&${v}_{\pm1}(\mathcal I,x_{00})$      &$0.0425 $ &${\eta}_\pm(\mathcal I,x_{00})$& $0.55$       & $0.4$       \\ 
    &      &    &${0.13333}$ &${-0.1260}$&${0.0778}$ 
&${\mathcal M}_{BC2}^0$               &${0.0415}$& -                           &${0.533}$&${0.397}$  \\  \hline 
$6$ &$\pm1$&BC23&$0.26667$     &$0.10304$      & $0.11875$ 
&${v}_{\pm 1}(\mathcal I,x_{00})+0.02$&$0.0775$&${\eta'}_\pm(\mathcal I,x_{00})$&$1.$    & $0.35$        \\ 
    &      &    &${0.26667}$&${0.08737}$&${0.120}$ 
&${\mathcal M}_{BC2}^0{+0.0116}$  &${0.0830}$ & -                       &${1.07}$ &${0.361}$ \\ \hline \hline  
$10$& 0    &BC2 &$-0.44444$     &$-0.95345$    &$-2.10$ 
&$v_0(\mathcal I,x_{00})$           &$4.25$  &${\eta}_0(\mathcal I,x_{00})$& $-1.8$        &  $3.5$         \\
    &      &    &${-0.44444}$&${-1.0187}$&${-2.17}$ 
&${\mathcal M}_{BC2}^0$               &${3.86}$& -                           & ${-1.79}$ & ${5.47}$      \\ \hline  
$10$& 0    &BC23&$0$            &$-0.16325$     &$0.53$ 
&$v_0(\mathcal I,x_{00})$-2.2         & - & $-1.75$                                       & - & -     \\ 
    &      &    &${0}$       &${-0.2286}$& ${-0.589}$ 
&${\mathcal M}_{BC2}^0{-2.38}$    & - & ${1.59}$                               & - & - \\ \hline  
$10$&$\pm1$&BC2 &$0.22222$      &$-0.3063725$   &$0.3525$ 
&$v_{\pm 1}(\mathcal I,x_{00})$       & $0.305$ & $ {\eta}_\pm(\mathcal I,x_{00})$ & $0.95$ & $0.89$ \\
    &      &    &${0.22222}$ &${-0.3499}$ &${0.360}$ 
&${\mathcal M}_{BC2}^0$               &${0.320}$& - & ${0.889}$ & ${0.822}$  \\ \hline
$10$&$\pm1$&BC23&$0.44444$      &$0.286235$     &$0.548$ 
&$v_{\pm 1}(\mathcal I,x_{00})+0.001$ &$0.53$ &${\eta'}_\pm(\mathcal I,x_{00})$    & $1.75$ & $0.9$   \\
    &      &    &${0.44444}$ &${0.2427}$ &${0.558}$ 
&${\mathcal M}_{BC2}^0{-0.0472}$   &${0.640}$ & -          & ${1.78}$ & ${0.725}$ \\ \hline \hline
$20$& 0    &BC2 &$-0.88889$     &$-3.815$       &$-16.5$ 
&$v_0(\mathcal I,x_{00})$             &$80.$ & $ {\eta}_0(\mathcal I,x_{00})$& $-3.5$ & 45.     \\
    &      &    &${-0.88889}$&${-4.075}$  &${-17.4}$ 
&${\mathcal M}_{BC2}^0$               &${61.7}$ & - & ${-3.56}$ & ${82.6}$ \\ \hline  
$20$& 0    &BC23& $0$           &$-0.653$       &$-4.0$ 
&$v_0(\mathcal I,x_{00})$-28.         & - & $20.$ & - & -  \\
    &      &    &${0}$       &${-0.914}$ &${-4.71}$ 
&${\mathcal M}_{BC2}^0{-27.8}$    & -    & ${20.4}$ & - & - \\ \hline  
$20$&$\pm1$&BC2 &$0.44444$      &$-1.22536$     &$2.804$ 
&$v_{\pm 1}(\mathcal I,x_{00})$       &$4.475$ &${\eta}_\pm(\mathcal I,x_{00}$)    & 2.335 & $11.6$ \\
    &      &    &${0.44444}$ &${-1.400}$  &${2.88}$ 
&${\mathcal M}_{BC2}^0$               &${5.12}$ & -         & ${1.78}$ & ${8.16}$ \\ \hline 
$20$&$\pm1$&BC23&$0.88889$      &$1.14545$      &$4.323$ 
&$v_{\pm1}(\mathcal I,x_{00})-0.875$& $7.0$ & $ {\eta'}_\pm(\mathcal I,x_{00})$& $3.5$ & $10.5$ \\ 
    &      &    &${0.88889}$ &${0.9707}$&${4.46}$ 
&${\mathcal M}_{BC2}^0{-1.47}$& ${10.2}$ & - & ${3.56}$ & ${20.7}$ \\\hline
  \end{tabular}
	\normalsize
  \caption{\label{tab:fits} 
The table presents the first coefficients of the expansion of $\mathcal M(x_{max})$ in powers of 1/$x_{max}$, obtained by a fit performed with $x_{max}$ spanning the interval $[20$~ru$,500$~ru$]$.
 The non-resonant light intensity $\mathcal I$, the magnetic quantum number $m$, and the asymptotic boundary conditions BC are specified in columns 1, 2 and 3, respectively. The intensity-dependence of the coefficients is indicated below the coefficients in the second line of the top cells.
 The numerical calculations of $\mathcal M(x_{max})$ include three coupled channels $\ell = 1$, $3$, $5$ using 
 $150$ or $200$ values of the nodal parameter $x_{00}$, chosen such that the field-free $s$-wave scattering length varies from $-\infty$ to $+\infty$. For given $\mathcal I$, $m$ and BC, and for each value of $x_{00}$, the calculated function $\mathcal M(x_{max})$ is fitted to the analytic expansions~\eqref{eq:fit}. In each cell of cell-lines 2 to 13 and columns 4 to 11, the first line reports either the numerical values of the coefficients of the terms, when it does not vary with $x_{00}$, or the symbol $v(\mathcal I,x_{00})$ or $\eta(\mathcal I,x_{00})$ for the $x_{00}$-dependent ones. The second line gives the analytical results obtained with the Levy-Keller formulas~\ref{eq:MI2} and~\ref{eq:MI23}, for the single-channel $\ell=1$ and for the potential $V^m_{ad}(x)$~\autoref{tab:pot}. Except for the BC23 reference pair and $m=0$, the coefficients $ {v}_m(\mathcal I,x_{00})$ and $ {\eta}_m(\mathcal I,x_{00})$ are related by a linear transformation $ {\eta}(\mathcal I,x_{00})= {v}(\mathcal I,x_{00})\times {\mathrm \alpha} - {\mathrm \beta}$. The values of ${\mathrm \alpha}$ and ${\mathrm \beta}$ (which do not depend on $x_{00}$) are reported in the first line of the cells of columns 10 and 11. The second line of these cells reports the  factor $\mathcal M_0$ and the constant term in the analytic expression of the $1/x$ coefficient 
   in~\autoref{eq:MI2} and~\autoref{eq:MI23}. } 
\end{table*}
%
%

The fitting coefficients of  $\mathcal M(x)$ for the terms $x^2$, $x$, $\ln(x)$, $1$, $\ln(x)/x$ and $1/x$, see~\autoref{eq:fit}, and obtained from numerical multi-channel calculations are presented in~\autoref{tab:fits}. The analytical ones obtained with the Levy-Keller approach using the asymptotic effective potential $V^m_{nad}(x)$  in the $p$-wave, see~\autoref{tab:pot}, are also presented. For fixed $m$, $\mathcal I$ and boundary conditions BC,  the numerical and analytical coefficients are shown in the upper and lower lines, respectively, of the same cell in~\autoref{tab:fits}.

%

We encounter coefficients independent of $x_{00}$, their numerical values are specified in~\autoref{tab:fits}, whereas others depend on $x_{00}$. 
The coefficients of the constant and $1/x$ terms, labeled by the symbols $ {v}_m(\mathcal I,x_{00})$  and  ${\eta}_m(\mathcal I,x_{00})$, respectively, depend on $x_{00}$ with a shape presenting several divergences (see Fig.~\ref{fig:scatt-dim}).  Note that for BC23 and $m=0$, ${\eta}_m(\mathcal I,x_{00})$ is independent of $x_{00}$, and its value is given in~\autoref{tab:fits}.

For given $m$ and $\mathcal I$, the $ {v}_m(\mathcal I,x_{00})$ of the BC2 and BC23  boundary conditions  differ by a constant, which is  independent of $x_{00}$ and is listed in the first line of a BC23 cell in column 7. The corresponding analytical difference is expressed in terms of $c_3$ and $c_{3f}$ and  is reported 
in the second line of the same cell.  
The multichannel numerical values agree well with the estimates obtained in the single-channel approximation 
in~\autoref{eq:M0CI2-M0CI23}. 
Similarly, for given $m$ and $\mathcal I$, the coefficients ${v}_m(\mathcal I,x_{00})$ and ${\eta}_m(\mathcal I,x_{00})$ are related by a linear transformation ${\eta}(\mathcal I,x_{00})= {v}(\mathcal I,x_{00}) {\mathrm \alpha}-{\mathrm \beta}$, with the ${\mathrm \alpha}$ and ${\mathrm \beta}$ coefficients reported in cells of columns 10 and 11 of~\autoref{tab:fits}. Here, we also find a good agreement between the fitted and single-channel approximation results reported 
in the same cell in the upper and lower lines, respectively.

The intensity-dependence of the coefficients in the $\mathcal M(x_{max})$ expansion are obtained from the single-channel formulas and the expansion of $V_{ad}^m(x)$. This dependence, indicated in the second line of  the top cells  of~\autoref{tab:fits}, is reproduced by the numerical fits. At low intensity, the $\mathcal I/x^2_{max}$ contribution prevails (column 4), whereas for increasing intensity the contribution of higher orders such as $\mathcal I^4 \times \ln(x_{max})/x_{max}$ (column 8) becomes important. The difference $\mathcal M^0_{BC23} - \mathcal M^0_{BC2}$ varies as $a \mathcal I^3 +b \ln(\mathcal I)$  (column 7). For BC32 and $m=0$, the coefficient of $1/x_{max}$ varies as $1/3 +b' \mathcal I^4$ (column 9), the first term arising from the van der Waals interaction. The same dependence occurs for the factor $\beta$ (column 11).

The numerical values obtained by fitting the multi-channel ($n=3$) results agree well with the single-channel ($p$-wave) approximation coefficients derived using an adiabatic potential, cf. upper and lower lines in each cell of~\autoref{tab:fits}. In particular, both calculations reproduce the classification of the coefficients into two types: The first one characteristic of the asymptotic $p$-wave potential, the other one accounting for the interactions at short distances. In addition, we emphasize that the values of $\mathcal M_0$ corresponding to the BC2 and BC23 boundary conditions are equivalent and  are related by a general expression  depending  only on the asymptotic potential, see~\autoref{eq:M0CI2-M0CI23}. 

The results from~\autoref{tab:fits} justify the extraction from the multi-channel calculations, or more precisely from the expansion of the divergent $\mathcal M(x_{max})$ into powers of $1/x_{max}$, a term independent of $x_{max}$, $ {v}_m(\mathcal I,x_{00})$. This quantity plays the same role as $\mathcal M_0$ in the single-channel approximation. Thus, as in the single-channel approximation, we introduce in the multi-channel model a generalized scattering volume given by ${v}_m(\mathcal I,x_{00})$, which characterizes low-energy collisions when the dynamics is governed by an anisotropic $1/x^3$ interaction. 
%
\bibliography{shaperes}
\end{document}